\documentclass[a4paper,twoside,10pt]{article}

\def\s{\sig\mua}

\def\b{\bigskip}
\def\bb{\bigskip\bigskip}

\def\no{\noindent}
\def\r{\rightline}
\def\ce{\centerline}
\def\ve{\vfill\eject}

\def\r{\rightline}
 
 \def\g{{\got g}}
\def\L{{\cal L}}

 \def\s{\sig\mua}
\def\\mueti#1{\par\indent\llap{#1\enspace}\ignorespaces}
\def\harr#1#2{\s\muash{\\muathop{\hbox to .25 in{\rightarrowfill}}
 \li\muits^{\scriptstyle#1}_{\scriptstyle#2}}}

\def\diagra\mu#1{{\nor\muallineskip=8pt
 \nor\mualbaselineskip=0pt \\muatrix{#1}}}

\def\R{{\cal R}}

\def\today{\ifcase\month\or January\or February\or March\or April\or
May\or June\or July\or
August\or September\or October\or November\or  December\fi
\space\number\day, \number\year }

\r \today

\def\DD{\vec \bigtriangledown}
\def\Rr\mu{\hbox{\rm I\hskip -2pt R}}
\def\Nr\mu{\hbox{\rm I\hskip -2pt N}}
\def\Cr\mu{\hskip0.5\mu\mu \hbox{\rm l\hskip -4.5pt C\/}}
\def\w{\wedge}

\def\e{\rm e}

\def\p{\partial}

\def\sqr#1#2{{\vcenter{\vbox{\hrule height.#2pt
\hbox{\vrule width.#2pt height#2pt \kern#2pt
\vrule width.#2pt}
\hrule height.#2pt}}}}

 \def\square{\mathchoice\sqr34\sqr34\sqr{2.1}3\sqr{1.5}3}

 \def\1/2{{\scriptstyle{1\over 2}}}
 \def\a/2{{\scriptstyle{3\over 2}}}
 \def\5/2{{\scriptstyle{5\over 2}}}
 \def\7/2{{\scriptstyle{7\over 2}}}
 \def\3/4{{\scriptstyle{3\over 4}}}

\font\steptwo=cmb10 scaled\magstep2

\def\picture #1 by #2 (#3){
  \vbox to #2{
    \hrule width #1 height 0pt depth 0pt
    \vfill
    \special{picture #3} % this is the low-level interface
    }
  }

\def\scaledpicture #1 by #2 (#3 scaled #4){{
  \di\muen0=#1 \di\muen1=#2
  \divide\di\muen0 by 1000 altiply\di\muen0 by #4
  \divide\di\muen1 by 1000 altiply\di\muen1 by #4
  \picture \di\muen0 by \di\muen1 (#3 scaled #4)}
  }
\input epsf

\def\sqr#1#2{{\vcenter{\vbox{\hrule height.#2pt
\hbox{\vrule width.#2pt height#2pt \kern#2pt
\vrule width.#2pt}
\hrule height.#2pt}}}}

 \def\square{\mathchoice\sqr34\sqr34\sqr{2.1}3\sqr{1.5}3}

\def\g{\Gamma}

\def\s{\sig\mua}

\def\b{\bigskip}
\def\bb{\bigskip\bigskip}

\def\no{\noindent}
\def\r{\rightline}
\def\ce{\centerline}
\def\ve{\vfill\eject}

\def\r{\rightline}
 
 \def\g{{\got g}}
\def\L{{\cal L}}

 \def\s{\sig\mua}
\def\\mueti#1{\par\indent\llap{#1\enspace}\ignorespaces}
\def\harr#1#2{\s\muash{\\muathop{\hbox to .25 in{\rightarrowfill}}
 \li\muits^{\scriptstyle#1}_{\scriptstyle#2}}}

\def\diagra\mu#1{{\nor\muallineskip=8pt
 \nor\mualbaselineskip=0pt \\muatrix{#1}}}

\def\R{{\cal R}}

\def\today{\ifcase\month\or January\or February\or March\or April\or
May\or June\or July\or
August\or September\or October\or November\or  December\fi
\space\number\day, \number\year }

\def\DD{\vec \bigtriangledown}
\def\Rr\mu{\hbox{\rm I\hskip -2pt R}}
\def\Nr\mu{\hbox{\rm I\hskip -2pt N}}
\def\Cr\mu{\hskip0.5\mu\mu \hbox{\rm l\hskip -4.5pt C\/}}
\def\w{\wedge}

\def\e{\rm e}

\def\p{\partial}

\def\sqr#1#2{{\vcenter{\vbox{\hrule height.#2pt
\hbox{\vrule width.#2pt height#2pt \kern#2pt
\vrule width.#2pt}
\hrule height.#2pt}}}}

 \def\square{\mathchoice\sqr34\sqr34\sqr{2.1}3\sqr{1.5}3}

 \def\1/2{{\scriptstyle{1\over 2}}}
 \def\a/2{{\scriptstyle{3\over 2}}}
 \def\5/2{{\scriptstyle{5\over 2}}}
 \def\7/2{{\scriptstyle{7\over 2}}}
 \def\3/4{{\scriptstyle{3\over 4}}}

\font\steptwo=cmb10 scaled\magstep2

\def\picture #1 by #2 (#3){
  \vbox to #2{
    \hrule width #1 height 0pt depth 0pt
    \vfill
    \special{picture #3} % this is the low-level interface
    }
  }

\def\scaledpicture #1 by #2 (#3 scaled #4){{
  \di\muen0=#1 \di\muen1=#2
  \divide\di\muen0 by 1000 altiply\di\muen0 by #4
  \divide\di\muen1 by 1000 altiply\di\muen1 by #4
  \picture \di\muen0 by \di\muen1 (#3 scaled #4)}
  }

\def\sqr#1#2{{\vcenter{\vbox{\hrule height.#2pt
\hbox{\vrule width.#2pt height#2pt \kern#2pt
\vrule width.#2pt}
\hrule height.#2pt}}}}

 \def\square{\mathchoice\sqr34\sqr34\sqr{2.1}3\sqr{1.5}3}

\def\g{\Gamma}

\usepackage{graphicx, float, mwe}

\def \X{\dot{\vec X}}
\def\P{\DD\Phi}
\def\N{{\cal N}}
\bb
\begin{document}

\ce{\steptwo Rotating planets in Newtonian gravity}
 
%Seliger

\hskip2in  \bb

\ce{\it Christian Fronsdal} 

\bb

\small {\ce{ \it Dep.t of Physics and Astronomy, University of California Los Angeles, CA, USA}}

nnn
aaa
 
{\it ABSTRACT}                    Variational techniques have  {\bf been used in applications of 
hydrodynamics in special cases. What is needed is an Action that is general enough to deal with potential flows as well as  vortex flows, with rotating fluids in Nature and in the laboratory; it has  become available only recently.  This paper is one of several that aim to test and develop a new Action Principle for  hydrodynamics. Here we study models of rotating planets,  compressible fluid  bodies in a stationary state of motion, under the influence of a fixed gravitational field. The hope is to account for the shape and the flow velocities, given the size of the equatorial bulges, the angular velocity at the equator and the density profiles.  The theory is applied to the principal objects in the solar system, from Earth and Mars, to Saturn with its famous hexagonal flow and its characteristic ring system. Planetary rings are an unforeseen but, as it it turns out, a natural and  inevitable feature of the dynamics;  past cataclysmic events  are not needed to explain their existence.  Observed  stresses may have been created during a period of quasi-static changes as orbital angular momentum was transformed into spin.  

 This paper is preparation for a systematic application of a new action principle to a detailed study  of the  planets.   The present intention is to test the versatility of the action principle in astrophysical applications, while meeting some objections that can be raised against traditional methods.} 
 \b

fronsdal@physics.ucla.edu   \hskip1.8in fronsdal.physics.ucla.edu
\b

\epsfxsize.5\vsize
%\centerline{\epsfbox{3dimring.eps}}
 \parindent=1pc

\ve

\ve

\no{\bf  \Large 1. Introduction}

\ce {\bf 1.1. Classical hydrodynamics} 

\bf { {\b In this paper a classical approach to hydrodynamics is used to determine the ideal shape of a rotating heavenly body under the influence of its own gravitational field. The development of an effective action principle for hydrodynamics is now close to realization.  Applications in   different fields are  examined  in order to gain a wide perspective.  We begin with a brief summary of recent developments of the basic concepts.

Classical hydrodynamics comes in two `versions'.  
The most popular one, the `Eulerian version'; is defined by the action that was discovered by Lagrange (1760) [1],
$$
A_1 = \int dtd^3x\L_1,
~~~~\L_1 = \rho(\dot \Phi  -  \DD\Phi^2/2 -\varphi) - W[\rho].\eqno(1.1)
$$
In this theory the velocity is irrotational $\vec v := - \DD\Phi$. The scalar field  $\varphi$ is the Newtonian potential. \footnote {The inclusion of gravitation in this equation has been derived from Einstein's equation for the metric.}  The term $W[\rho]$ is the input from thermodynamics; it will be shown that, for an isolated system, it can be identified with the internal energy density.

This theory has a normal canonical structure with one pair of canonical variables,
the density $\rho$ and the velocity potential $\Phi$.  The equations of motion are
$$
{\delta A_1\over \delta \Phi} = \dot\rho + \DD\cdot(\rho\vec v) = 0 \eqno(1.2)
$$ 
and the Bernoulli equation (in integrated form)  
$$
{\delta A_1\over \delta \rho}  = \dot\Phi - \DD\Phi^2/2 -\varphi - 
{\delta W\over \delta \rho} = 0.
$$
The gradient of this equation is the  Bernoulli equation 
$$
\DD(\dot\Phi - \DD\Phi^2/2 -\varphi) = \DD {\p W[\rho]\over \p \rho} =  
{1\over \rho} \DD p;  \eqno(1.3)
$$
justified as follow.  The thermodynamic internal energy density is $f(\rho, T)  + sT $; $f$ is the free energy density and $s$ is the entropy density. The basic thermodynamic equations are 
$$
{\p f\over \p T} + s = 0,~~~~ {\p f\over \p \rho}-\rho{\p f\over \p \rho} + p   = 0.\eqno(1.4)    
$$
The first equation is used to eliminate the temperature; then the  internal energy density $u(\rho,s) = f+sT$ can be identified with $W[\rho]$.  
 
  Under the assumption that {\bf the entropy density $s$  is a linear function of the density, $s = \rho S$, with $S$ uniform,} it follows that 
$$
\DD {\p (f+sT)\over \p\rho} = {1\over \rho}\DD p, \eqno(1.5)  
$$ 
where $p := \rho\p f/ \p \rho - f$. Comparison of (1.3) with (1.5) shows that $W[\rho]$ is identified with $u(\rho,s) =  f + st$.
  In view of this result the Bernoulli equation takes the original form (Bernoulli 1743) [2],
$$
\DD(\dot\Phi - \DD\Phi^2/2 -\varphi)  = 
{1\over \rho}\DD p. \eqno(1.6) 
$$
{\bf Remark.}  This derivation of the Bernoulli equation may not be well known.   It is important to the author because it elevates  a set if equations to a theory with known limitations and conservation laws, with well defined hamiltonian, angular momentum and kinematic potential.    It is also important that Eq.(1.6) can be justified only if the specific entropy density $S$ is uniform; this limits its applicability, a fact that is only occasionally recognized.  A hydrodynamic  equation of state is an expression for $u(\rho,S)$ or $p(\rho,S)$. Solving the hydrodynamic equations of motion will give us these functions for one  value of $S$.  
\b 
 
 Lagrange's action principle has remained  popular, in spite of notorious difficulties, some of them summarized by the d'Alembert paradox of 1747 [3]; it states that flight is impossible.   A more elementary example is the popular experiment that consists of placing a glass with water on a turntable and observing the shape of the surface. This would be explained by the centrifugal force that is the negative gradient of the kinematic potential, $-\DD \Phi^2/2$, except that it appears in (1.6) with the wrong sign! The significance of this has been emphasized by the author (2020) [4].

 This is why a second  classical theory has been mustered, an `altenative version'
 of hydrodynamics, usually  credited to Lagrange. Here the velocity is not a gradient, but a time derivative; we shall denote it $\X$. The Lagrangian density of this theory is
$$
\L_2 = \rho\X^2/2  - W[\rho],\eqno(1.7)
$$
It dates from the same period as $\L_1$.  In this case the kinematical potential has the right sign; this theory is always used to `explain' the turntable experiment, in spite of the fact that this theory is defective by not including an equation of continuity. The Newtonian potential can be included here as well, but it can not be derived from Einstein's equation.

Thus neither version of the classical theory stands up to scrutiny.

The most elementary, stationary solutions of the two theories are circular flows of two kinds
$$
\DD\Phi = {a\over r^2} (-y,x, 0), ~~~~\X = b(-y,x,0).~~~~a,b ~~{\rm const.}\eqno(1.8)
$$
Cylindrical Couette flow has always been modelled by a linear combination of both, but the idea that the two theories should be combined in an action principle has been slow to emerge. The first step in this direction was taken by Landau in 1941 [5].  

%Of attempts to merge the two theories, by far the most famous is the theory
%of Phonons and Rotons proposed by Landau (1941). Later several authors invented actions for Landau's theory, but the ones I know of  used two copies of (1.6); this in spite of the fact that one of the velocities, 

The phonon field $\vec v_s$ of superfluids is always assumed to be (locally) irrotational.
The idea that the other velocity,  the roton field $\vec v_n$, is governed by $\L_2$ appeared in 1964, in an important paper by Rasetti and Regge [6].

A familiar analogy will help us appreciate this contribution. Consider the state of electromagnetism before Maxwell managed to combine electricity and magnetism, a theory of two vector fields with 6 variables.  Maxwell's theory was quickly organized around the vector potential and three of the equations were recognized as constraints, solved by setting 
$$
F_{\mu\nu} = \p_\mu A_\nu - \p_\nu A_\mu.
$$
Of the remaining  equations, one is a gauge theory constraint,
which reduces the number of physical degrees of freedom to 2. With the new focus on Lorentz invariance and the demands of unitarity, it did not take  long before the first gauge theory was born.  
\b

The great merit of the paper by Rasetti and Regge is that they accomplished a parallell development for the field $\vec X$. The result has found important applications in string theory.  (Kalb and Ramond 1974 [7],  Zheltukin 2014) [8].  It is a central part of the 2-vector, conservative form of hydrodynamics that is being applied and tested in the present paper.

The result of Rasetti and Regge is an imaginative (but unique) solution to this problem, a solution that embodies all the new aspects of Landau's theory of superfluids.
The unique, unitary, relativistic theory that solves the riddle is the 2-form relativistic gauge theory developed by Ogievetski and Polubarinov (1964) [9]. It has only one propagating mode and even this one is  irrelevant in most  non-relativistic applications.   
\b

Seliger and Whitman (1968) [10] made another  serious effort to discover an action that would combine the two versions of classical hydrodynamics,
in a theory that would have a nondegenerate canonical structure without constraints.  Their theory   has found few applications and it allows for vorticity only if the specific entropy {\bf is not} uniform.
\b

The relativistic  theory has an antisymmetric tensor field with components $Y_{\mu\nu}$, with six components.  The three components $\eta_i = Y_{0i}$ can be transformed to zero by a gauge transformation and there remains only the 3-vector
 $$ 
 \X^i = {1\over 2}\epsilon^{ijk}Y_{jk}.
 $$
 The Lagrangian density of the relativistic gauge theory, with the extra density factor, is $\rho(dY)^2$ where $(dY)_{\mu\nu,\lambda}$ is the antisymmetric differential of $(dY)_{\mu\nu}$.  
The field equations that come from variations of the gauge field give rise to the constraint
$$
\DD\w \vec m = 0,~~~~\vec m := \rho(\X + \kappa \DD\Phi),~~~~\eqno(1.9)
$$
It is solved by setting
$$
\vec m =  - \DD\tau,\eqno
$$
where $\tau$ is a so far arbitrary scalar field. 
\b

 {\bf Remark.} In the papers on relativistic field theories that have been cited, as in many papers on hydrodynamics, the density was taken to be uniform, which prevents a useful contact with non-relativistic physics.
 \b

{\b The unique propagating mode in  this theory is   $\N := \rho(\DD\cdot \vec X+\kappa).$ It is massless ($\square \N = 0$) and  it is non-propagating  in the context of the present paper, 
$$
\Delta \N = \ddot {\N} = 0. \eqno(1.10).
$$

We arrive at a simple Lagrangian:
$$
\L = \rho(\beta\dot \Phi - \Phi^2/2 -\varphi)   + \rho\X^2/2 + 
\rho\kappa \X\cdot\DD \Phi   -W[\rho],~~~~\beta = 1+\kappa^2.\eqno(1.11)
$$
The  action  principle  incorporates traditional hydrodynamics,  including the equation of continuity and the Bermoulli equation with a velocity potential that is the sum of the two contributions, and a spin-orbit  interaction that is essential, as we shall show now.

The equation of continuity (see Eq.(1.2)) is  
$$
\dot\rho +\DD\cdot(\rho\DD\vec v) = 0,~~~~ \vec v := \kappa \X - \DD \Phi.\eqno(1.12)
$$
The vorticity is thus
$$
\DD\w \vec v = \DD\w {\kappa\over \rho} \vec m  = -\DD\w({\kappa\over \rho}\DD\tau).
 $$
 The constant $\kappa$ must  be non-zero, otherwise the new theory has only irrotational flows.
\b

\no{\bf Remark.} The last equation shows that the vorticity is perpendicular to the density gradient. Has this been confirmed by observation?
\b

{\b The theory has been extensively applied; to the  stability of cylindrical Couette flow (Fronsdal 2020b) [11], to capillarity and metastable states of thermodynamic fluids,  and recently to Gravitational Waves (Fronsdal 2021) [12] and to calculate the speeds if the sounds in  superfluid Helium. It provides the first field - theoretic model of rotating, relativistic  fluids that respects the Bianchi identity and the equation of continuity. %[The special case $\rho = 1$ is unphysical.]   

The principal feature of this paper is the action principle; in other respects it does not go as far as earlier models of planetary dynamics.
See for example  Beauvalet, Lainey, Arlot and Binzel (2013)  [13] or Stute, Kley and Mignone (2013) [14]. A feature that is not always included in planetary dynamics is the requirement that the mass flow velocity field be harmonic, as it needs to be in all stationary or quasi-static  models. (See below.) Most important, this restriction reduces the number of adjustable parameters and this increases the value of the action principle as a framework with a greatly improved predictive power.  
 
 \bb

\ce{\bf   1.2. Summary. The non relativistic model}

      The gauged-fixed, non-relativistic  matter action 
$$
\L = \rho(\beta\dot \Phi - \Phi^2/2 -\varphi)   + \X^2/2 + 
\kappa \X\cdot\DD\Phi   -W[\rho],\eqno(1.13)
$$
where  $\beta := 1+\kappa^2$,  is composed  of the Lagrangians of the two classical theories.
The variables are the density $\rho$ and the two velocity  potentials, 
the scalar  velocity potential $\Phi$ and the vector potential $\vec X$.   (Compare Schutz 1970 [15].)  The $\kappa$ term is needed; without it the mass flow becomes irrotational, as we have shown. The kinematic  potential has terms of either sign ($- \Phi^2/2$ and $+ \X^2/2$), as needed for different, elementary  applications. The relativistic relativistic origin  of the irrotational part (1.1) is in Fronsdal (2007) [16].
\b 

 The relation of the gauge fixed Lagrangian (1.11) to the complete set of field equations is as follow.  1. The gauge field $\eta$ has been fixed, so the main constraint  (1.9) is  a typical gauge theory constraint.  2. The condition (1.10) excludes travelling waves in the non-relativistic  sector.   3.  The equation of continuity,
$$
\dot\rho + \DD\cdot(\rho\vec v) = 0,~~~ \beta\vec v :=  \kappa\X  - \DD\Phi,\eqno(1.14)
$$
 comes from variation of (1.11) with respect to $\Phi$.    4. Variation of  $\vec X$ gives 
$$
{d\over d t}\vec m = 0. \eqno(1.15)
$$
 5. Finally, variation of the Lagrangian density with respect to the density gives the Bernoulli equation in the form
$$
\DD(\beta\dot\Phi - K -\varphi)  = 
{1\over \rho}\DD p,  \eqno(1.16) 
$$
where $K$ is the kinematic potential.
$$
K := -\X^2/2 -\kappa\X\cdot\DD\Phi + \DD\Phi^2/2.
$$
The theory is a combination of the two classical theories (`phonons' and `rotons'). The kinetic potential has terms of both signs, as needed in elementary applications. The negative sign of the first term means that it must be interpreted as stress.
  
 The on-shell value of the 
Lagrangian density is the thermodynamic pressure; as prophesized with rare insight by Taub (1954) [17].
\b

 The incorporation of two velocity fields is an essential feature of the theory, and of hydrodynamics:
what is both novel and effective is that they give it the minimal number of dynamical variables: four including the density. The Lagrangian has one free parameter $\kappa$; it  is inversely related to the  compressibility of the fluid.

 The  action is not completely new; parts of it appeared in a classical paper by  Hall and Vinen (1956) [18] on superfluids  and in a more recent review by Fetter (2009) [19] on rotating Bose-Einstein condensates.\footnote  {The equations of motion in the paper by Hall and Vinen are widely quoted; the action principle has been   completely ignored.} 
In those papers $\dot{\vec X}$ is not a local, dynamical field variable but a fixed  background feature that accounts for a rigid rotation  of the whole system. The dynamical,  irrotational velocity was clarly insufficient
and another degree of freedom was needed, but the way to avoid  an excessive number of new  degrees of freedom by means of constraints was not widely known.   
\bb

A stationary flow is one that evades the dissipating effect of {\bf viscosity}. In traditional hydrodynamics viscosity is included as an additional term in the Navier-Stokes equation,
$$
\dot{\vec v} + (\vec v\cdot\DD)\vec v = {-1\over \rho}\DD p + \bar\mu\rho\Delta 
\vec v,\eqno(1.17)
$$
where $\bar\mu$ is the kinematic viscosity. Viscosity normally implies dissipation and 
cannot be accommodated within an action principle, but its effect can be
acknowledged by  replacing the conservation law (1.15) by
$$
{d\over dt} (\rho \vec w) = \bar\mu\rho\Delta \vec v.
$$
In this way a theory based on conservation laws can be distorted to include a type  of dissipation, as   in the familiar approach with Eq.(1.16). In both theories,   stationary motion is possible only when the field $\vec v$ is harmonic. 
\bb

 \ce{\bf 1.3.  Planetary  rings}
 
   The biggest surprise to emerge from this work is the  spontaneous appearance of  planetary rings, of the type observed on minor planets.  Solitary rings are predicted, not just confirmed; they are prominent features of the simplest solutions of the equations of motion. 
 
The first planetary ring was found on the minor planet Chiron  in 1993. (The minor planet was discovered in 1977.)  Haumea's less spectacular ring  was first  seen in 2004. The third and last  of 
these rings  was observed on Chariclo on 26 March 2014. The first version of this paper was completed in December 2017 and submitted to arXiv.org on 20 May 2018, too late to claim a real prediction. (Source Wikipedia)   

{\bf These are the only isolated planetary rings found so far in the Solar system. 
The rings were discovered as a highly unexpected byproduct of a calculation of planetary shapes. It is perhaps the most spectacular confirmation of the aptitude of  Conservative Hydrodynamics and the Lagrangian (1.10) so far.} 

%\epsfxsize.5\vsize
%\centerline{\epsfbox{3dimring.eps}}4.
% \par-indent=1pc
% 
% 

 %%%%%%%%%%
% FIGURES 1 & 2 %
 %%%%%%%%%%
  
\begin{figure}[H]
    \centering
    \begin{minipage}{0.45\textwidth}
        \centering
        \includegraphics[width=0.2\vsize]{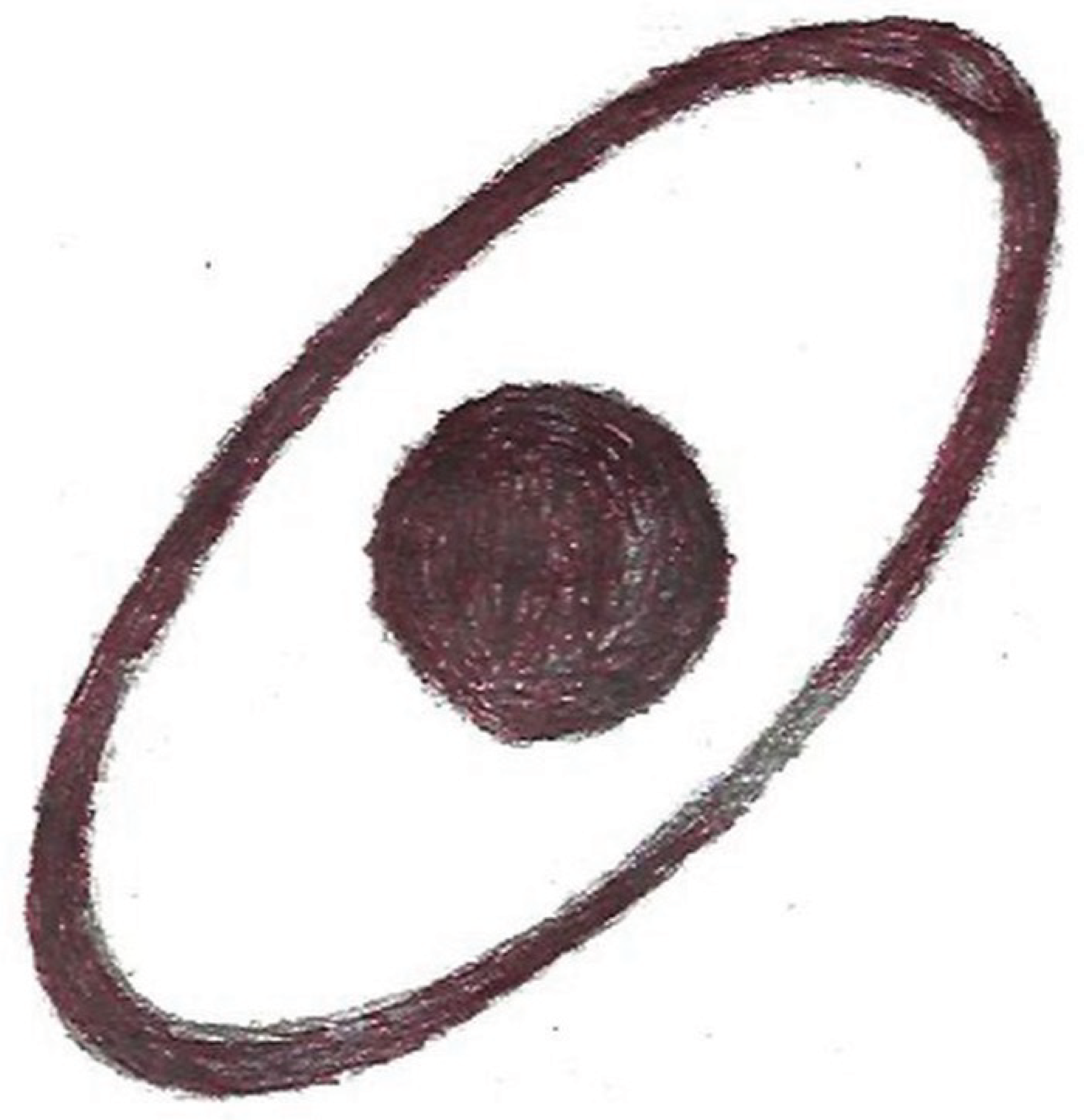} 
        \caption{One of the first rings that appeared, unexpectedly, in this investigation.}
    \end{minipage}\hfill
    \begin{minipage}{0.45\textwidth}
        \centering
        \includegraphics[width=0.2\vsize]{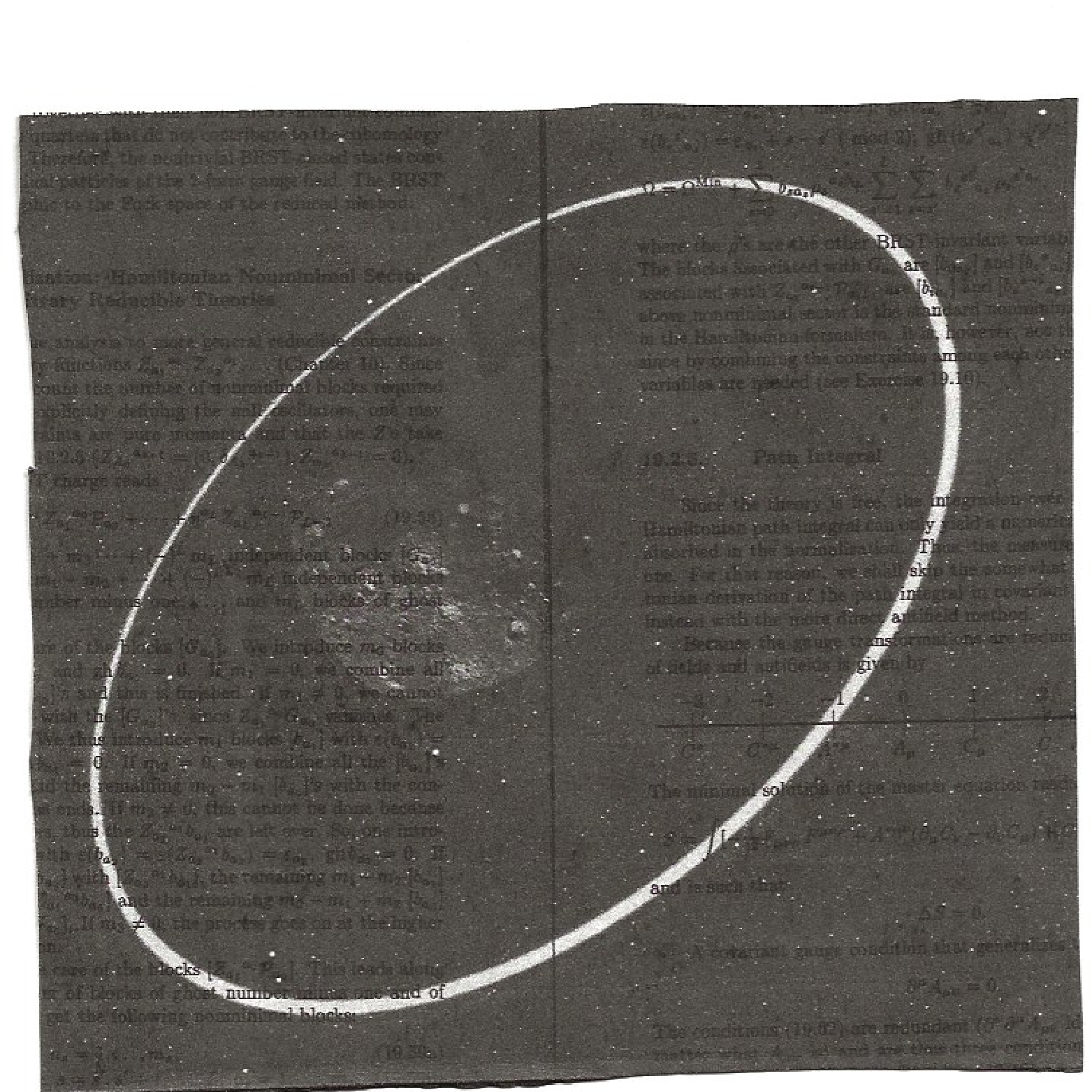}
        \caption{the First ring observed on the minor planet Chivron, in 1993.}
    \end{minipage}
\end{figure}

%\vskip0cm
% \hskip-5.5cm
%\epsfxsize.2\vsize
%\centerline{\epsfbox{Fig1.eps}}
%\parindent=1pc
%
%
% 
% \vskip-4cm
%\hskip2cm
% \epsfxsize.2 
%\vsize
%  \centerline{\epsfbox
%{Fig1.eps}}
%\vskip0cm
%
%
%\bb
%
%
%\small{Fig.1. One of the first rings that appeared, unexpectedly, in this investigation.}
%
%\small{Fig.2. the First ring observed on the minor planet Chivron, in 1993.}
%\b 

It is suggested that the parameter $N$ in Eq.(2.9) is related to evolution, that all the planets may have had rings at one time, that the planet Mars, in particular, may have had  a ring at relatively recent times  and that all the planets will eventually loose their rings.

   Extended ring systems, such as seen on the Jovian planets (Jupiter, Saturn, Uranus and Neptune), are located very close to the equator. This suggests a flow pattern with  high angular momentum. Harmonic flows concentrated on the equatorial plane are described in terms of  the gradients of separated  harmonic functions 
 $$
 \DD J_0(kr)\sinh(kz),
 $$
 where $r$ and $z$ are  cylindrical coordinatest. See Fig 16.
 
 The dimensional parameter $1/k$ determines the order of magnitude of the average spacing between the rings as well as their  thickness; but the actual spacing is irregular. 
 \b

 {\bf Remark.} Chiron is said to have ``rings like Saturn".  This suggests that 
 the latter may be susceptible of modelling without much more effort than those of Chiron and Chariclo.
\bb

\ce {\bf  1.4. The Cylindrical Couette Problem}

{\b The most familiar systems treated in Newtonian gravity as well as in General Relativity have spherical symmetry. But most heavenly bodies  are rotating
around an axis that is more or less fixed, with approximate cylindrical symmetry. Much of the inspiration for our work comes from a laboratory  experiment with similar properties,  cylindrical Couette flow. Taking the advice of Homer Lane (1870) [20], as is traditional in astrophysics, we apply to astrophysics what we have learned in terrestrial laboratories. The problem examined by Couette (1888-1890) [21] and Mallock (1888, 1896) [22-23]  
is a fluid confined between two concentric cylinders that can be rotated
independently around the vertical axis. In classical hydrodynamics the balance of forces is expressed by the Navier-Stokes equation. Boundary conditions are  assumed to be non-slip and the fluid is compressible. At low speeds any stationary motion has been  described by the following  vector field
$$
\vec v = {a\over r^2}(-y,x,0) + b(-y,x,0),~~~r := \sqrt{x^2+y^2},~~~a,b~{\rm constant}.\eqno(1.18)
$$
 The first term is irrotational for $r \neq $ 0 and both are harmonic.   
\b

The new action principle was used to account for the stability of basic, cylindrical Couette flow.  %One feature of the model is that the density profile is subjected to a strong condition that originates in the demand that $\vec v$ be harmonic.

The study of Couette flow has revealed one aspect of the physical interpretation: it is responsible for fluid stress and some metastable configurations, and it is indispensable for the analysis of capillary action and other surprising properties of water.  
\ve

\no{ \bf  \Large 2. A compressible fluid rotating in a fixed, central} 

\ce{\bf \Large gravitational field}

Our model of a planet is an isolated system with a liquid or solid core in thermodynamic equilibrium with a gaseous atmosphere, in a stationary, rotating state and described by the Lagrangian (1.11). It consists of a single substance in two phases. The condition of thermodynamic equilibrium at the phase boundary is that the pressure $p$, the temperature and the chemical potential $\mu$ be continuous across the surface. In the case of a thin atmosphere this implies that the pressure and the chemical potential are constant on the surface.  The surface is thus a locus of the function $K + \varphi$; see Eq.(1.16).

$$
C(\vec x) := \dot{\vec X}^2/2 + \kappa \X\cdot\DD\Phi - \DD\Phi^2/2 -{-GM\over R},\eqno(2.1)
$$
$$
R = {r\over |\sin\theta|} = \sqrt{x^2+y^2+z^2}. 
 $$
We are using the Newtonian approximation for the attractive gravitational potential. It is an expedient shortcut  of the present treatment, as in the simplest version of the traditional approach, and one that we hope to remove later.  It may be a valid approximation so long as the departure from spherical symmetry is small.

This model should be appropriate for Earth and Mars and possibly for the frozen planets Neptune and Uranus,  less so for the gaseous planets. To determine the appropriate velocity fields we begin by examining the simplest solutions.
\b

 If  the velocity is irrotational, and $\dot{\vec X}$ = 0; then the shape is determined by 
$$
C_1(\vec x) = -{a^2\over 2}{1\over r^2} +{GM\over R} = {\rm constant},
$$
$$
r := R|\sin\theta| = \sqrt{x^2+y^2}.
$$

A plot of the loci of this expression for several values of the  parameter $a^2$ reveals, instead of an equatorial bulge, a polar depression. This attempt, postulating an irrotational flow, evidently fails. 
\vskip0cm

 Solid-body flow   
is the complementary case in which  $\DD\Phi = 0$ and  the angular velocity $\omega = b$ is a constant. As in the traditional approach; the condition of equilibrium is 
 
$$
C_2(\vec x) = {\omega^2\over 2}r^2 + {GM\over R}  =  ~{\rm constant}.
$$
There is a bulge.  The number usually quoted is
$$
\epsilon := {R_{eq.}\over R_{pole}}-1 = {\omega^2\over 2MG}R_{eq.}^3,
$$
where, to a good approximation, $R_{eq}$ can be replaced by 1 on the right hand side. For Planet Earth the number is
$
MG = R^2g,~ g = 980 cm/sec^2,
$
and the approximate value of $\epsilon$ is predicted by this model to be
$$
{\omega^2R^3\over 2MG} = {R\omega^2\over 2g}
=
\left({2\pi\over 24\times 3600}\right)^2{6.357\times10^8\over 2\times 980} = .0017.\eqno(2.2)
$$
Solving Eq. (2.1) for the azimuthal angle we find that
$(b^2/ 2) R^2\sin^2\theta = C - {GM/ R}$ and we conclude that, with
the solid body hypothesis
 
\b

a. the radius is minimal at the poles  and  

b. the locus of (3.1) always has a non-compact branch. 
\b

The observed value of $\epsilon$ for Earth is .00335,  twice the prediction (2.2). The classical theory can be improved by taking into account the effect of the bulge on the potential; for example, by assuming that the shape is an ellipsoid, and that the density is uniform. That results in a value for $\epsilon$ of .0042, which is too large (Fitzpatrick 2018) [24]. %Ref    
A further improvement results from taking the partly known density distribution into account; this has the effect of diminishing the effect of the shape on the potential. 

\b

\no{\bf Remark.}  This is the classical result, but the traditional calculation does not get it from an action principle; instead it postulates a kinetic potential 
$-\vec v^2/2 = -(\omega r)^2/2$. In this case  the Bernoulli equation is {\it ad hoc} and the continuity equation cannot be justified. %When we use (1.4) in the special case wuen $\Phi = 0$ we get $\vec v = 0$ from (1.2). 
This disregard for rigor is commonly repeated, even  in General Relativity. It is especially deplorable 
in an educational setting when rigour yields to simplicity.
\bb

 \ce{\bf  2.1 The general case, 2 flows}

{\bf We look for the general solution of the equations of motion. Again 
$$
\varphi = -{GM\over R},~~~\R := \sqrt{x^2+y^2+z^2},~~~~G = ~{\rm constant}. \eqno(2.3)
$$
The full set of equations includes the equation of continuity and 
$$
\vec v := \kappa \X - \P,~~~~\Delta \vec v = 0,~~~\vec w: = \X + \kappa\P = {-1\over \rho}\DD\tau,\eqno(2.4)
$$
%$$
%\dot\rho +\DD\cdot(\rho\vec v) = 0,~~~{d\over dt}(\rho\vec w) = \mu\rho\Delta \vec v, 
%$$
$$
 \X^2/2 + \kappa\X\cdot\P - \P^2/2 + {GM\over R} =\mu[\rho],\eqno(2.5)
$$

%To have a stationary motion we must either suppose that  viscosity is zero or else limit the motion by  including the condition that the vector field $\vec v$ is harmonic. In this we are imitating the classical analysis of Couette flow. %In an attempt to preserve generality we look for solutions with the property  that $\Delta\vec v = 0$. 
 
All the vector fields can be expressed in terms of the two scalar fields $\Phi$ and $\tau$  and the density.  The simplest possibility is that the two flows are  in planes perpendicular to the axis of rotation, 
$$
-\DD\Phi = {a\over r^2}(-y,x,0),~~~-\DD\tau = {b\over r^2}(-y,x,0).
\eqno(2.6)
 $$  
These are  gradient-type vector fields  with angular momentum $L_z = \pm 1$. They would not be sufficient for an ambitious attempt to construct realistic models, but they may be enough for our main purpose,  to establish the versatility of the action principle. Other solutons   will be used to account for the extraordinary  ring systems of the Jovian Planets.
\b

  Eq.s (2.4) and (2.6)  give the velocity of mass flow
$$
\vec v = \kappa\vec w - (\kappa^2+1)\DD\Phi = \omega(-y,x,0),\eqno(2.7) 
$$
with the angular velocity
$$
\omega := {1\over r^2}\left({\kappa b\over \rho}  + a(\kappa^2+1)\right).\eqno(2.8)
$$
The most general harmonic vector field of this form is a series of spherical functions with higher  angular momenta.   There is  evidence of higher
angular momenta in the flow velocities  of Venus, Pluto and, most notably, Saturn, to be discussed later. Here we reduce the series to the simplest terms, the normalized inverse density taking the form
$$ 
{1\over \rho} = 1 + N R +\eta r^2 + \nu{r^2\over R^3},~~~~N > 0,~~~~\eta > 0,\eqno(2.9)
$$
with constant coefficients $N,\eta, \nu$. This type of solution is strongly indicated in the case of the Earth and the other planets with a rigid surface. The first two terms have $\ell = -1$, %\footnote { The introduction of a non-integrable representation of the rotation group  should be noted.}
 the others $\ell = 1$. Higher harmonics are needed in the case of Saturn, since this planet shows a distinct, hexagonal flow pattern. %The ring systems of Saturn and the other Jovian planets  are described by a different kind of harmonics. 
 
A non-zero value of the  last term in (2.9) would give rise to a hole, shaped like a donut, near the center of the planet; see Fig.3. It may serve as a regularizing device, but it is hardly relevant for the evaluation of the shape of the surface.    We  adopt  the expression (2.9), with $\nu = 0$, as a plausible first approximation to the density profiles of Earth and Mars and, very tentatively, to those of the other planets.

%%%%%%%%%%%
% FIGURE THREE %
%%%%%%%%%%%

\begin{figure}[H]
\centering
\includegraphics[width=0.8\hsize]{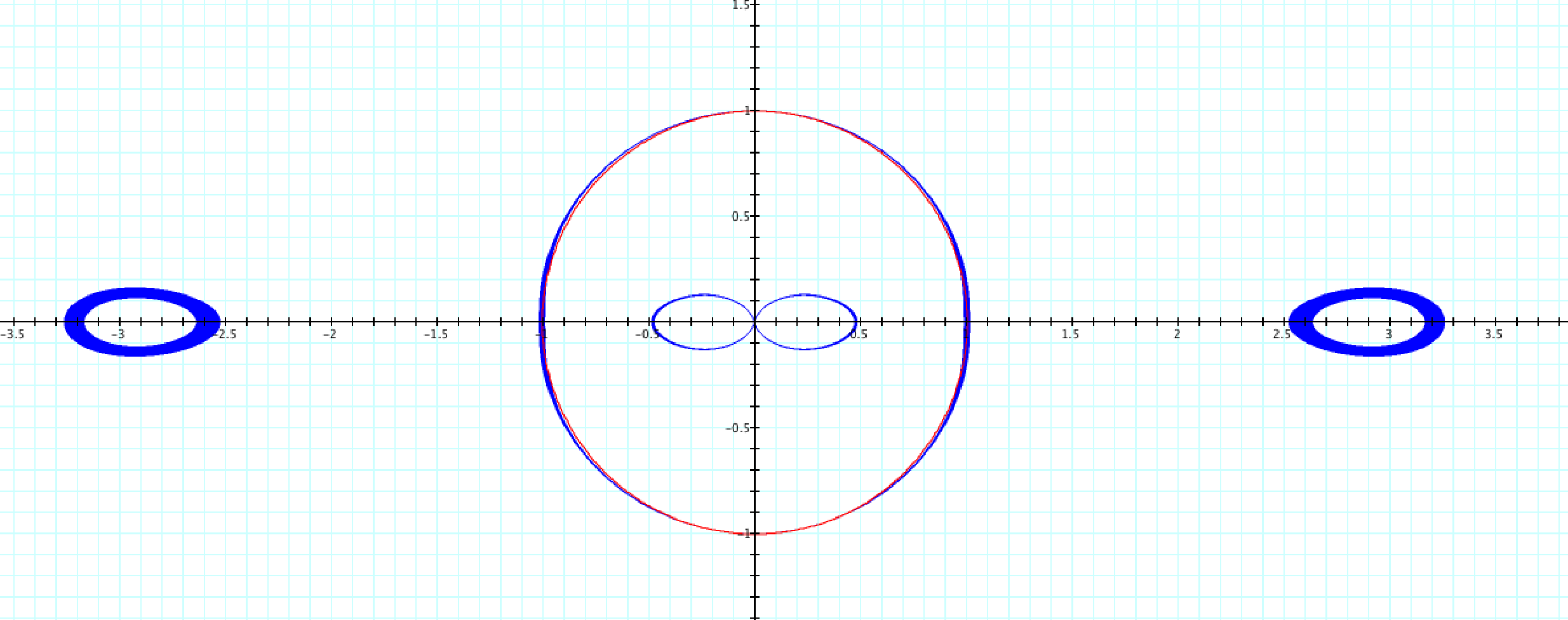}
\caption{The effect of including the 4'th term in (2.9); a hole appears at the center, as well as a ring in the equatorial plane. Shown here is a locus of the function $C$ in Eq.(2.1).}
\end{figure}

%\vskip0cm
%\vskip0cm
%\epsfxsize.8\hsize
%\centerline{\epsfbox{Fig.3.eps}}
%
%\
% 
%\hskip-3cm
%\vskip0cm
% \hskip-1cm
%   \epsfxsize.5\vsize 
%   \centerline{\epsfbox{Fig3.eps}}
% 
% 
%
%{\small  Fig. 3. The effect of including the 4'th term in (2.9); a hole appears at the center, as well as a ring in the equatorial plane. Shown here is a locus of the function $C$ in Eq.(2.1).}  
% \b

 Eq.s (2.8-9), with $\nu = 0$, is our simplest  model planet. It is intended, in the first place, to serve as a model for Earth and Mars and perhaps for Uranus and Neptune. A more detailed model  would  need a general harmonic expansion for the flow vector field $\vec v $.
 \b

 \ce{\bf  2.2. Units}
 
 The unit of density is the central density. The unit of length is the polar radius.
   
 \ve
 
\ce{\bf 2.3. The flow }

The flow  is very close to that of a solid body. Combining (2.8) and (2.9) with $\nu = 0$ gives
$$
\omega := {1\over r^2}\left(\kappa b(1+NR + \eta{r^2})  + a(\kappa^2+1)\right).\eqno(2.10)
$$
This becomes regular on the axis of rotation ($r=0$)  if
$$
a(\ \kappa^2+1) = -\kappa b (1+N).\eqno(2.11)
$$
This makes $\omega$ constant and is therefore strongly indicated for Earth.

That $a$ and $b$ have opposite signs implies that the direction of rotation may change sign within the planet.%  or that two planets with similar histories may be seen as rotating in opposite direction. 
  
\b

\ce{\bf 2.3.  Regularity at the poles} 

 Take the polar radius to be the unit of length and let the central density be the unit of density. Both $N$ and $\eta$ have to be positive, as we shall see. 
\b

Let us begin with  a star that is spherically symmetric ($\eta = 0$) with polar radius 1 and  density ratio 
$$ 
{\rho_{center}\over\rho_{pole}} = N+1;
$$ 
then we allow for a modest violation of spherical symmetry by increasing the parameter $\eta$ from zero. 
\b

With Eq.s (2.1), (2.7) and (2.8) the equation $C(\vec x) =$ constant  for the surface takes the form, 
$$
 {1\over 2r^2}\bigg(b^2(1 + NR+\eta r^2)^2 - a^2(1+\kappa^2)\bigg) +{GM\over R} = {\rm constant}.
$$
To avoid getting a dip at the poles (from the denominator $r^2$)
we must have 
$$
a^2(1+\kappa^2) = b^2(N + 1)^2.\eqno(2.12)
$$
Then the equation takes the form
$$
 {b^2\over 2r^2}\bigg((1 + NR + \eta r^2)^2 -(N+1)^2\bigg) +{GM\over R} = {\rm constant}.\eqno(2.13)
$$
The choice (2.12), like (2.11), is also strongly favored for Earth. Both together implies that $\kappa >> 1$, which is consistent with the rigidity of a large part of the earth's crust.
\b

 Finally the shape is determined by
$$
f(R,r)  
  := {(1 +  NR + \eta r^2)^2 - (N+1)^2 \over r^2}  + {\xi\over R} = {\rm constant},\eqno(2.14)
  $$
  where $\xi$ is the constant
  $$
\xi =   {2GM\over b^2}\eqno(2.15)
$$
with the dimension of inverse length. For Earth the number is
$$
\xi = {2MG\over R^3}{R^4\over b^2}= {1.357\over 1980}10^8{R^4\over b^2} = 67986 ({R^2\over b})^2\eqno(2.16)
$$

Solutions of (2.14) extend to very large $R$ only if there is an effective cancellation between the terms of highest power, in $NR +\eta r^2$. If $\eta N$ is positive there can be no cancellation, at any azimuth; hence all the solutions are compact when  we pose
$$
\eta N \geq 0.
$$
We shall find that both parameters are positive.
\bb

\ce{\bf 2.5.  Angular velocity at equator}

The value of the visible angular velocity at equator is found from Eq.(2.10). To an accuracy that neglects terms of relative order $\epsilon/R$,
$$
R^2\omega = \kappa b(1+N+\eta) + a(\kappa^2+1).
$$
With (2.12),
$$
\omega = {b\over R^2}\bigg(\kappa(1+N+\eta)-\sqrt{\kappa^2+1}(N+1)\bigg),\eqno(2.17)
$$
where $R$ is the polar radius
and for$,\kappa >> 1,$
$$
\omega \approx {b\over R^2}\kappa\eta.\eqno(2.18)
$$

\bb

\e 
\ce{\bf  2.4.  Overview of results}

The two parameters $N$ and $\xi$ form a 2-space with the latter as abscissa; it divides into a lower region (roughly $N < 1$) where the planets have a ring, and a complementary upper region where they do not, separated by a ``ring-no-ring" boundary. See Fig.4, where three versions of this dividing line are shown, for $\epsilon = 1/300$ and with $\eta =.01$, .05 and .1 from high to low.  In the same figure we have shown nearly vertical lines of dots, a ``trajectory" for each of four planets. The coordinates of the dots on each of the planetary
trajectories  give a near-perfect fit to the measured ellipsoid of the respective planet.

 %%%%%%%%%%
% FIGURE FOUR %
 %%%%%%%%%%

\begin{figure}[H]
\centering
\includegraphics[width=0.7\hsize]{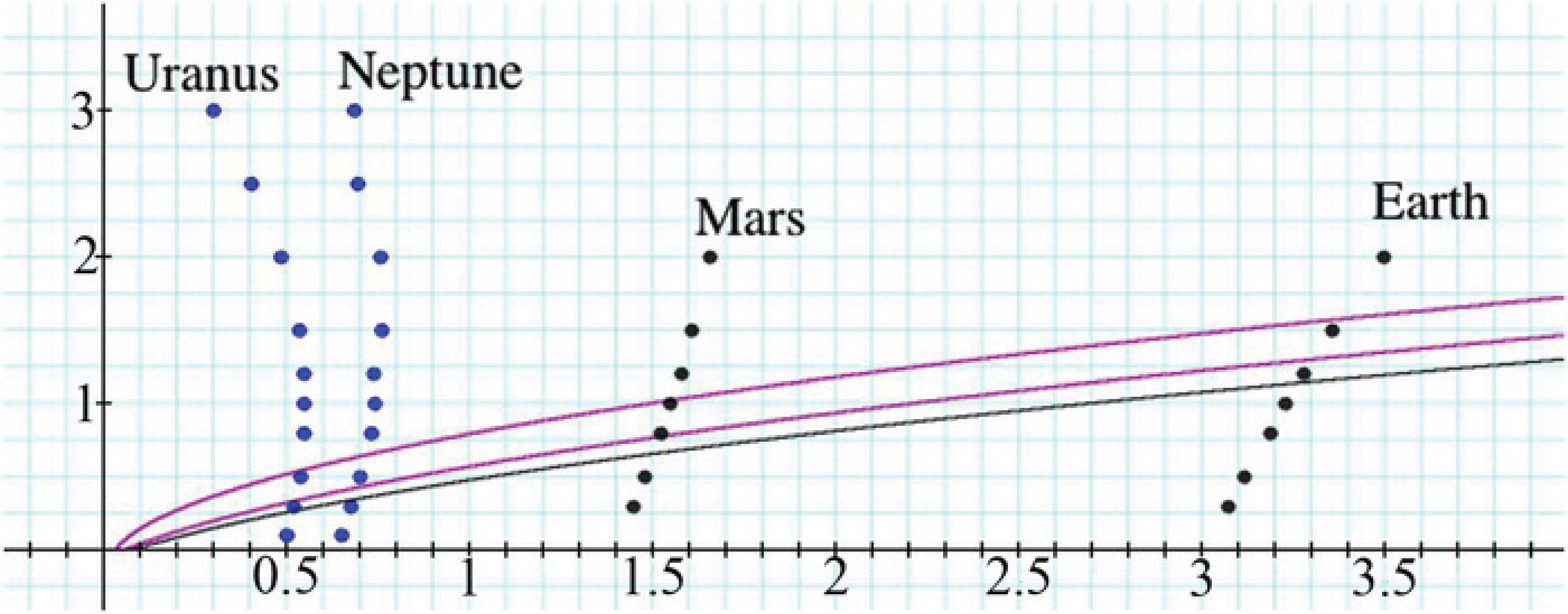}
\caption{The abscissa is the parameter $\xi$ and the ordinate is $N$.}
\end{figure}

%\b
%\epsfxsize.7\hsize
%\centerline{\epsfbox{Fig4.eps}}

Fig. 4 shows the result of calculations as points in the plane of the parameters $\xi$ and $N$, each dot representing a near perfect fit to the ellipsoid. Each line of dots  consist of points with coordinates that give perfect fits to the  shape of the respective planet, without rings on the upper part. Results of the calculation are tabulated in the Appendix.
\b
  
It may be permissible to think of this diagram as an {\bf evolution diagram}, each planet evolving upwards towards a state of greater  compression, and loosing its rings as it crosses the ring-no-ring dividing line. Earth  lost its rings long ago; it is to be placed on the upper part of its trajectory, well above the $\eta = .1$ line.
\b

Increasing $N$ means higher compressibility at the center;  Earth may have $N$ as high as 2 while  Mars is less compressed and may
have $N = 1$ or less. Since planets are likely to become more compressed over time we expect planets to evolve upwards. This is in accord with speculations that Mars may have had a ring in the evolutionary recent past. Uranus and Neptune still have rings and must have $N \leq .3$ if our model is applicable to them.
This means that they are  more compressible  than Mars and much more compressible than Earth, in agreement with observation. Results for Earth are listed in Table 1.

%Increasing $N$ is believed to mean increasing stiffness (reduceing compressibility); Earth may have $N$ as high as 2. Mars is less stiff and may 
%have $N = 1$ or less. Since planets are likely to become stiffer with time we expect planets to evolve upwards. This is in accord with speculations that Mars may have had a ring in the evolutionary recent past. Uranus and Neptune have rings and must have $N\leq .3$ if our model applies. % Leave this out? Pluto has lost any rings that it may have had and Neptune may be next in line to loose some.

  Other aspects of the model, including the equation of state, will be taken up in the connection with the  gaseous planets, in section 5.\
 \vskip-1cm
 
  \begin{table}[h]
\caption{Earth}
\begin{center}
\begin{tabular}{@{}cccccccccccc@{}}

  &    &    &   &  & &   &  & 
\cr
&Parameter & $N$ = .1 &.3 &  1 & 1.2 & 1.5 &  
\cr         
&  $\xi$  & 3.075   &   3.23  & 3.28 & 3.36 & 3.50  &   
\cr
&  $\xi_0$  & .6  &   2.7 & .28 &  4.9 &  7.8  &     
\cr
&  $\kappa\eta$  & 3.075/.6  &   3.23,2.7 & 3.28 & 3.36/4.9 & 3.50/7.8      & 
\cr

\hline
\end{tabular}
\end{center}
\end{table}

\bb

\no{\bf \Large 3.  Earth and Mars}

The polar radius is our unit of length. %The equatorial radius is determined by a 4'th order algebraic equation that is obtained from (3.1) by setting $r=R$.
We are mostly interested in shapes that are almost spherical (leaving aside the 
planetary rings for the moment), with a small equatorial bulge. To find surfaces that include a point  on the equator with radius $s=1+\epsilon$ we write  Eq.(2.14) in the form 
$$
B(R,r) :=   f(R,r) - f(s,s) = 0,\eqno(3.1)
$$
%Verify that this version was used.
Note that $f(s,s)$ is a constant.  The equatorial radius is a zero of the function $B(r,r)$ and after division by $r-s$ this equation reduces to a cubic. For planets without rings this cubic does not have positive roots. %If a ring is about to disappear at a distanse $s'$ from the center there will be a double zero at $r = s'$. %This method will be most effective later, when we are looking for planetary rings.
\b

The measurable parameter $\epsilon$ has replaced the value of the function $f$. However, there are still 3 parameters left, $N,\eta$ and $\xi$,  and it is difficult to survey all possibilities. We shall try to find  our way around this difficulty by looking at individual planets.
\b

If  Earth  is an ellipsoid  with eccentricity $\epsilon$  and the polar radius is normalized to unity, then the shape is
$$
R-1 =  \epsilon\sin^2\theta;~~~~R_{eq} = 1+\epsilon,
$$
with $\epsilon = .00335$. With $s = 1.00335$,  the locus $B(R,r) = 0$ 
passes through the equator at $R = r = s$, and through the pole. Fittings of shapes are relative to this ellipsoid, with the observed value of $\epsilon$.

 %Fig.2.4. The abscissa is the parameter $\xi$ and the ordinate is $N$.

%\epsfxsize.6\hsize
%\centerline{\epsfbox{Fig.2.5.Venus/Pluto.eps}}
\vskip.5cm

\parindent=1pc

%Fig. 2.5. Planets. The ring-noring lines are plotted as in Fig.2.4, but only the low est one is relevant; and that one has been plotted for the parameters of Earth. There is a different line for each planet but the difference is very slight. The message is: Uranus and Neptune, since they have rings, have very low values of $N$, less than .3, and Mars probably has $N \approx 0.7$.

\ce{\bf 3.1 The quality of the fits,  examples}

Very good fits to the ellipsoid are achieved with $N=2$ and both of the following  
$
 \eta = .092,~~\xi = 3,~~\eta = .1,~~\xi = 3.5.
$
The value $N=2$ was suggested by  the measured density profile 
shown below, in Fig.5.% and $\eta = .092$ is then a best fit.%; this value for the asymmetry parameter  seems too high. %The second solution was found by assuming a much smaller value for $\eta$; the resulting  value $N=2.8$ is reasonable. Both solutions were found by fitting the bulge parameter as well as the elliptical shape, both being sensitive to the parameters.

We tried $\eta = .1$ and $N=2$, leaving only $\xi$ to be varied. The locus is a curve that, at a small scale, resembles the geoid, the fit is perfect at the pole and at the equator.
We examined the error at nine intermediate azimuths and found that a perfect fit
would require $\xi$ to vary from 0 to 3. But if we fixed $\xi = 3$  the relative error  was never larger than $10^{-4}$.% We increased the value of $\eta$ to .1 and repeated the calculation. Now the value of $\xi$ required for a perfect match varied from 3 to 6 and when we fixed the value of
%$\xi$ at 3.5 the relative error was never larger than $10^{-5}$. %We take this as a hint that the value of $\eta$ may rise at least  as high as .1. 

The conclusion is that the identification of the planetary shape with a locus of $ C $ through the pole and the equator appears to be  natural and that the precise determination of the parameters applicable to each planet remains available for the fitting to additional data.  
\b

\ce{\bf 3.2 Density profile, range of $N$}

Earth is unique among the planets in that the density profile has been  reliably estimated, see Fig. 5.  A good fit to the central core is not possible since both $N$ and $\eta$ must be positive. This can be understood since the constitution of the earth is far from uniform; the model
assumption that the interior is a single phase is therefore an over-simplification. A fair approximation to the observed density suggests that $N$ lie in the interval
$$
1.5 < N < 2.5.\eqno(3.1)
$$

 %%%%%%%%%%
% FIGURES 5 & 6 %
 %%%%%%%%%%
  
\begin{figure}[H]
    \centering
    \begin{minipage}{0.45\textwidth}
        \centering
        \includegraphics[width=\hsize]{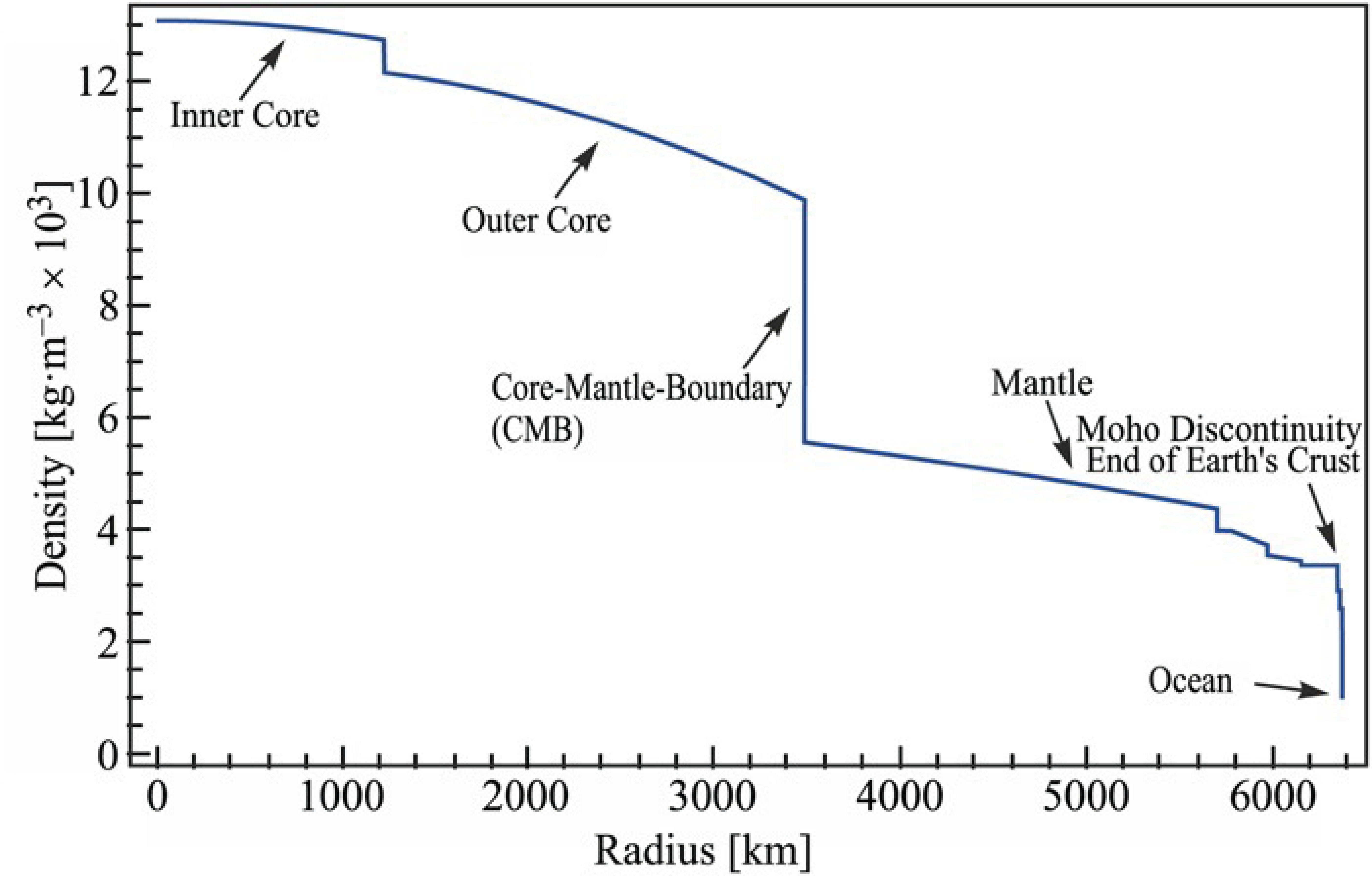} 
        \caption{Observationally estimated density profile of Earth.}
    \end{minipage}\hfill  
    \begin{minipage}{0.45\textwidth}
        \centering
        \includegraphics[width=\hsize]{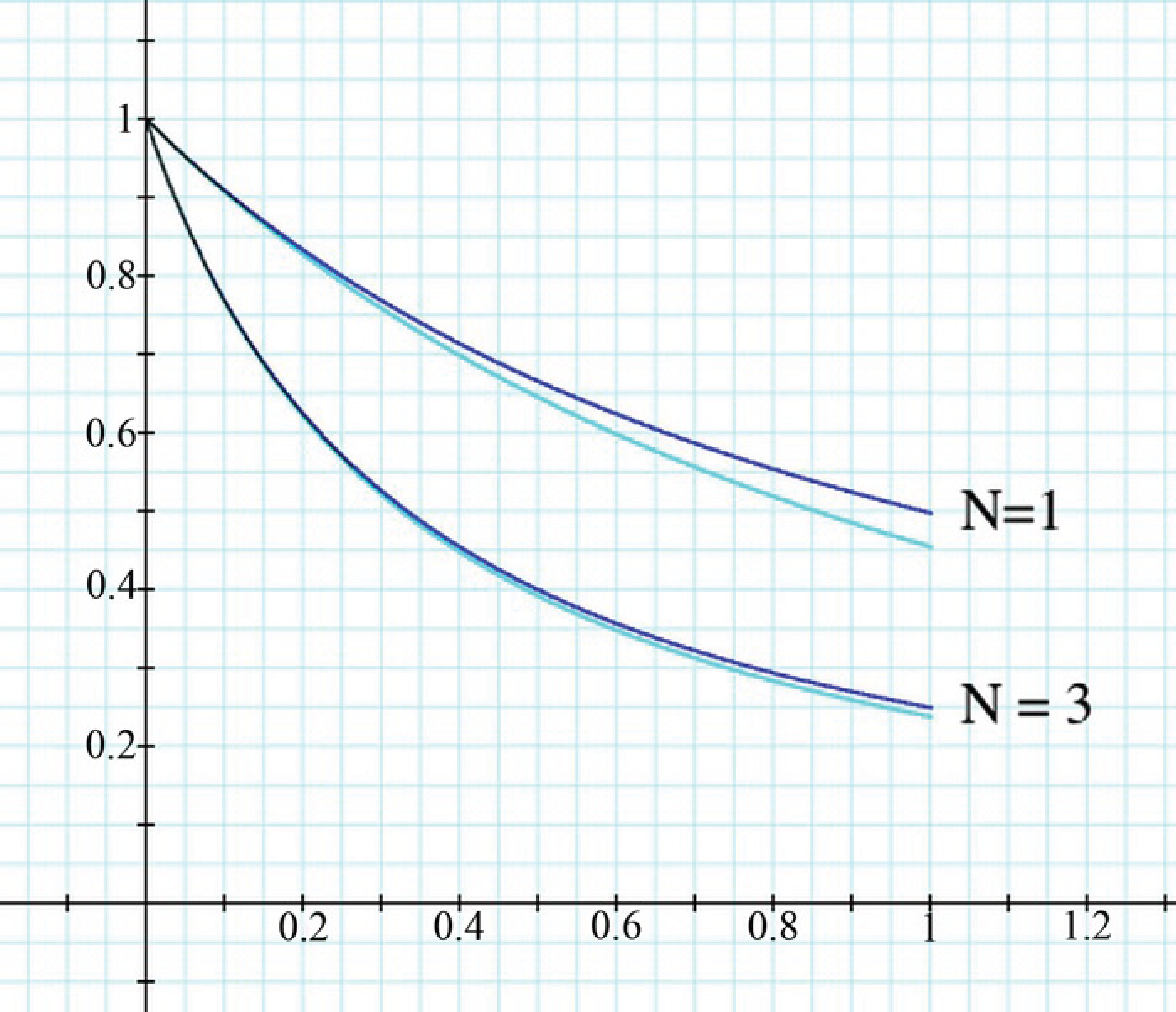}
        \caption{Model equatorial density profile of Earth for two values of the principal parameter $N$. Heavy lines $\eta = .01$, lighter lines, $\eta = .2$}
    \end{minipage}
\end{figure}

% 
% 
 %
% \epsfxsize.3\hsize
%\centerline{\epsfbox{Fig5.eps}}
%
% 
%    
%\vskip0cm
%\hskip0cm
% %%\epsfxsize.1\hsize
%\centerline{\epsfbox{Fig.6.eps}}
%
%\vskip-4cm
%
%{\small Fig. 5. Observationally estimated density profile of Earth.} 
%
%
%{\small Fig. 6. Model equatorial density profile of Earth for two values of the principal parameter $N$. Heavy lines $\eta = .01$, lighter lines, $\eta = .2$}.  
%\b
%\vskip0cm
%
 
\ce{\bf  3.3 Rings, or not}

Random sampling of the parameters of the theoretical configurations reveal that the expected, nearly spherical shape of the body is not always realized.
For example, in the case that $\eta = .1, N = 1.2$ we get a good approximation to the ellipsoidal shape of the Earth with $\xi = 3.3$.  But if the value of $\xi$ is increased  to 3.525, then a planetary ring appears, as shown in Fig. 4, moving horizontally towards the right.  For 
still larger values of $\xi$the ring
eventually dwarfs the planet.

\b

in Fig. 7 . We have crossed the ring-no-ring divider in Fig. 4, moving horizontally towards the right. For still larger values of $\xi$ the ring eventually dwarfs the planet.
\b
 
%%%%%%%%%%%
% FIGURE SEVEN %
%%%%%%%%%%%

\begin{figure}[H]
\centering
\includegraphics[width=0.5\hsize]{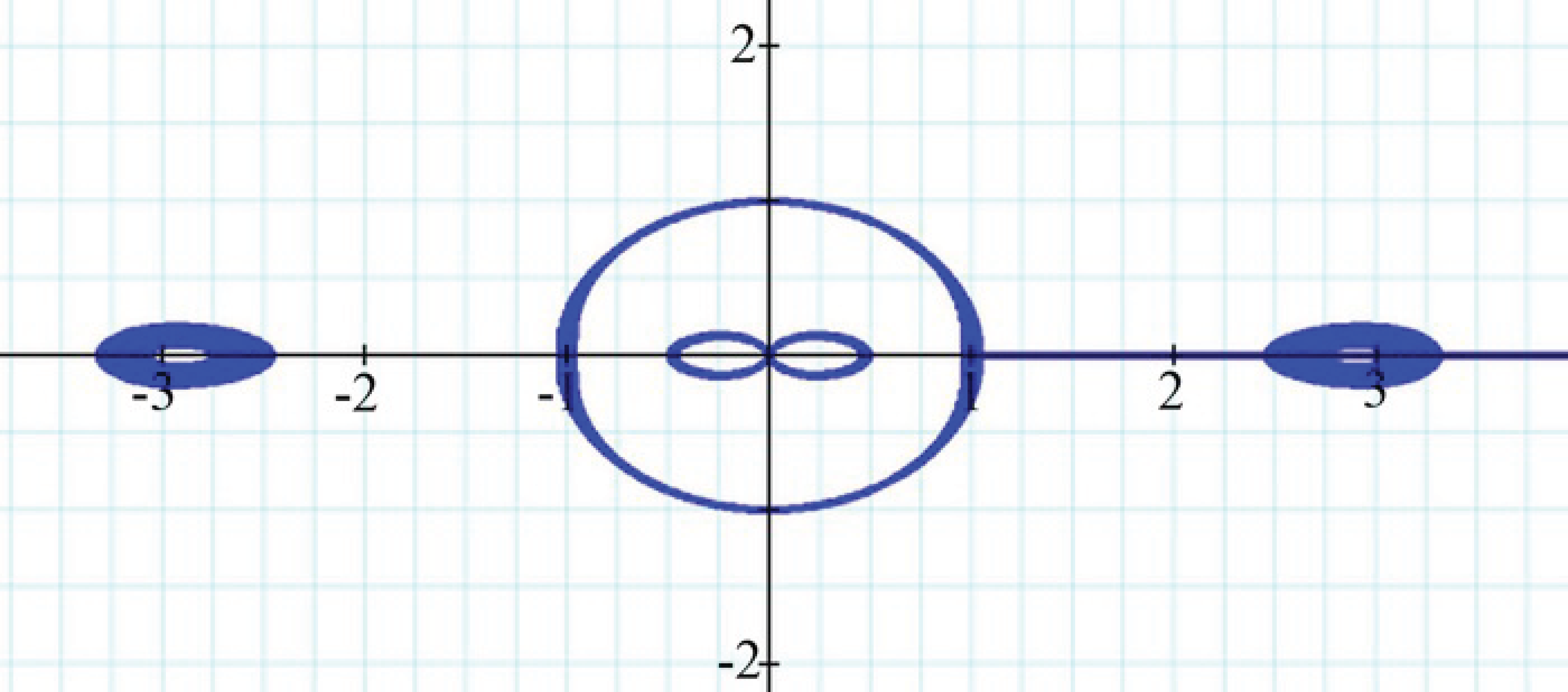}
\caption{Ring around Earth; about to disappear. Parameters $ N=1.2,  \xi = 3.525.$ For smaller values of $\xi$ there is no ring.}
\end{figure}
 
% 
% 
%\epsfxsize.5\hsize
%5\centerline{\epsfbox{Fig.7.earthRing.eps}}
%\vskip0cm
%\parindent=1pc
%
% 
% 
% 
%
% {\small Fig. 7. Ring around 
% Earthth; about to disappear. Parameters $ N=1.2,  \xi = 3.525.$
% For smaller values of $\xi$ there is no ring.}
% 
%  
%  

Nothing in our model relates to the hemispheric asymmetry of Mars; that is, the depression of the northern hemisphere. But we do know the bulge ratio, $\epsilon = 1/135$, and elaborate models of the planet  suggest a plausible density profile, shown in Fig. 5. We obtain excellent fits to the ellipsoid  from  $N=.7, \xi = 1.5$ to $N=3, \xi = 1.75$.

 %%%%%%%%%%
% FIGURES 8 & 9 %
 %%%%%%%%%%
  
\begin{figure}[H]
    \centering
    \begin{minipage}{0.45\textwidth}
        \centering
        \includegraphics[width=0.8\hsize]{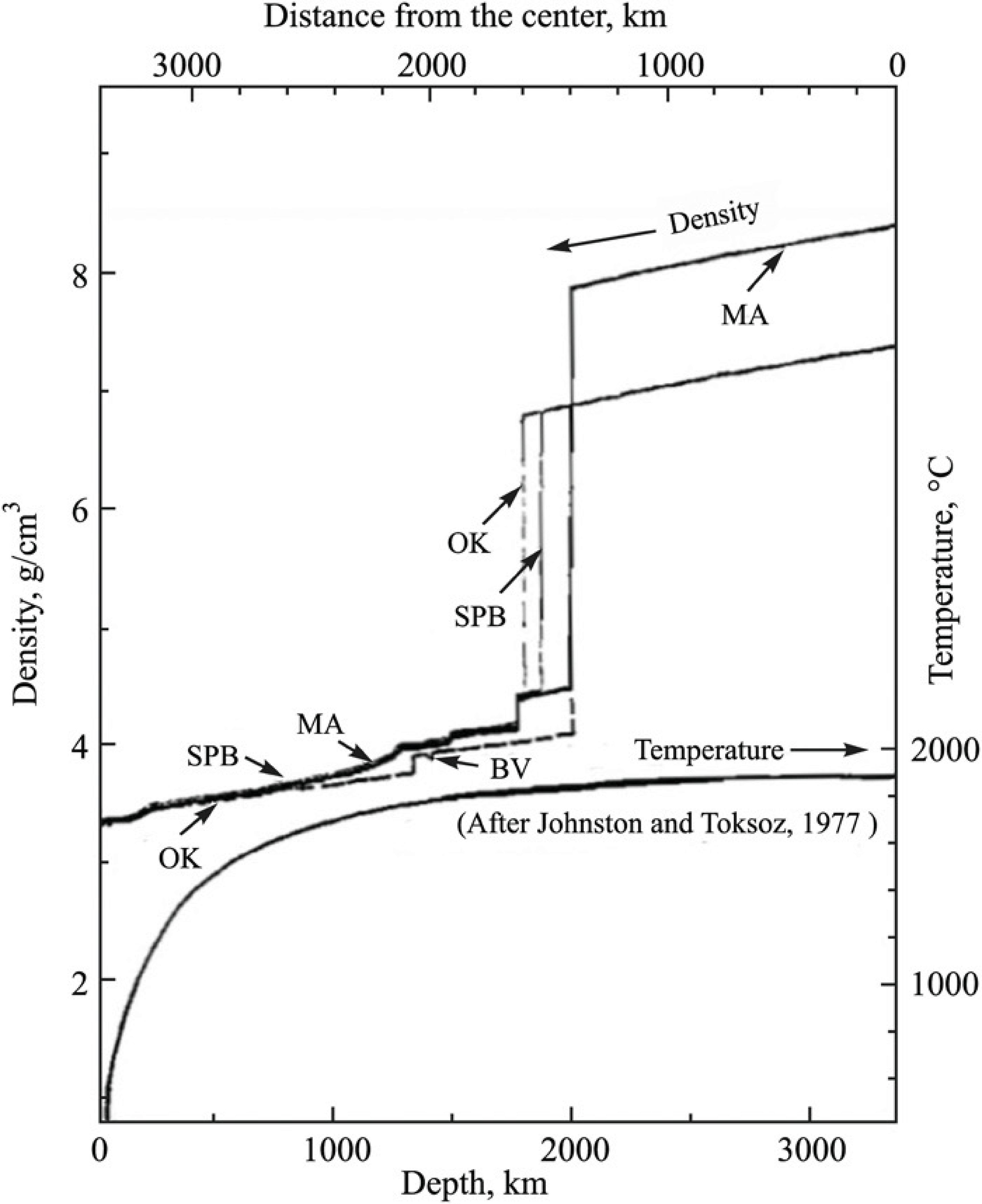} 
        \caption{Density profile of Mars has a thick mantle and a metallic core.}
    \end{minipage}\hfill  
    \begin{minipage}{0.45\textwidth}
        \centering
        \includegraphics[width=0.8\hsize]{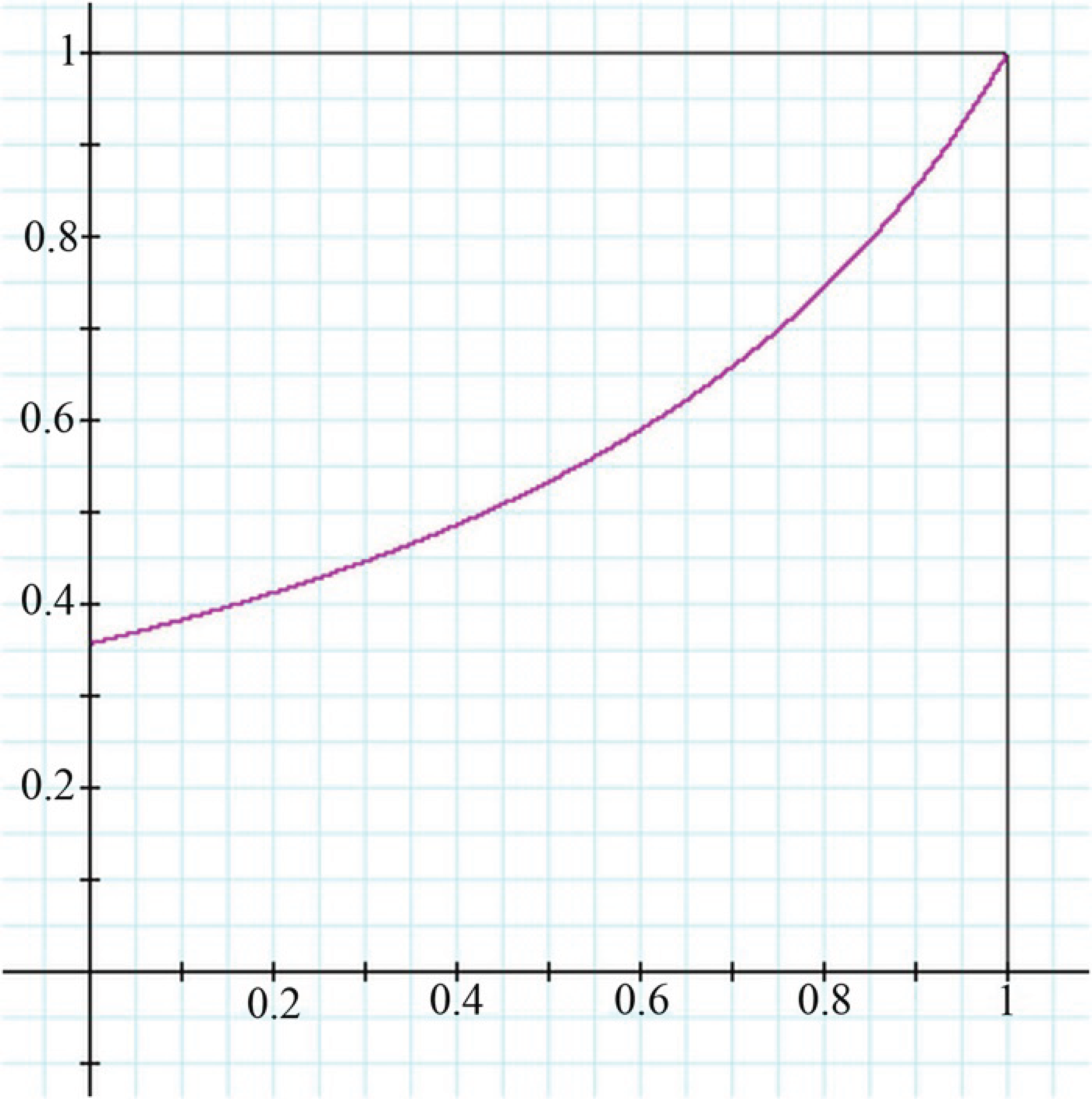}
        \caption{Density profile for Mars from Eq.(2.8), with $N$ = 1.7.}
    \end{minipage}
\end{figure}
  
%  \hskip-2.5cm
%\epsfxsize.2\hsize
%\centerline{\epsfbox{Fig6.eps}}\parindent=1pc
%
%
% 
%\hskip-2cm
% %\epsfxsize.25\hsize
%\centerline{\epsfbox{Fig.7.eps}}
%\vskip-2cm
% \parindent=1pc
% 
% 
% 
%
%   
%\vskip-1cm
%\hskip1cm
%\epsfxsize.2\hsize
%\centerline{\epsfbox{Fig.8'eps}}
% \parindent=1pc
% \vskip1.5cm
% 
%
% \end{doc}
%  
%{\small Fig. 8. Density profile of Mars  
%  has a thick mantle and a metallic core.}
%
% \vskip.3cm 
%
%{\small Fig. 9. Density profile for Mars from Eq.(2.8), with $N$ = 1.7.}
%\b
  
The idea of a ring around Earth can not be entertained, but the planet Mars is
another matter. The trajectory for Mars was calculated  with the equatorial bulge ratio $\epsilon = 1/135$. We fixed the  value   of 
$\eta $ at .1 as for Earth and for each of a sequence of values of $N$ we searched for the value of $\xi$ that gives the best approximation to the idealized shape of planet Mars.  The lowest value of $N$ for which such a fit exists is $N = .8$.
\b

It is notorious that Mars shows clear evidence of having once been furrowed
with large gulleys by the action of water (on the southern hemisphere). It has been widely  interpreted in terms of 
a cataclysmic event, eons past. Our calculations suggest that the value of $N$ was once lower, that a ring actually did exist around Mars, and that the ring   (consisting mostly of water or ice) disappeared as a result of the slow increase in $N$. The alternative, that the ring may have fallen as the result of a  passage very close to Earth, is less appealing, since we regard  the ring as  natural and  expect it to resume its original shape after a shock. If it did not, then it means that the property $N$ has evolved and that the equations of motion no longer supports a ring. The absence of surface features on the northern hemisphere also supports the idea that the fall of the ring was relatively non violent. For information about the Geology of mars see Carr and Head (2010) [25] and Carr, M. [26].
\bb
 
\no{\bf  \Large 4. Neptune and Uranus. Venus and Pluto}

  The shapes of Uranus and Neptune are quoted in the literature but actually they are poorly known (Bertka and Fei 1990) [27].  If we treat them as close analogues of Earth and Mars, then they would appear far to the left in 
  Fig. 4, as shown. The progression of values of the parameter $N$ from Neptune to Earth suggests increasing compression.
 
  \b

 Pluto is far out among the outer planets but its small mass is a more relevant parameter (Hellled, Anderson and Schubert 2010) [28].  Venus has a complicated structure and winds that are not  parallel to the equatorial plane. Both are essentially spherical but Venus has a small bulge ``that is  induced by winds". Venus also has a very dense atmosphere with a pressure of about 92 Earth atmospheres. For all these reasons neither Venus nor Pluto should be included in this study. Nevertheless,
we assigned very small bulges and got good fits to the ellipsoid  shape, some of which are recorded in Table 1 and plotted in Fig. 10. 

 %%%%%%%%%
% FIGURE TEN %
 %%%%%%%%% 

\begin{figure}[H]
    \centering
    \includegraphics[width=0.4\hsize]{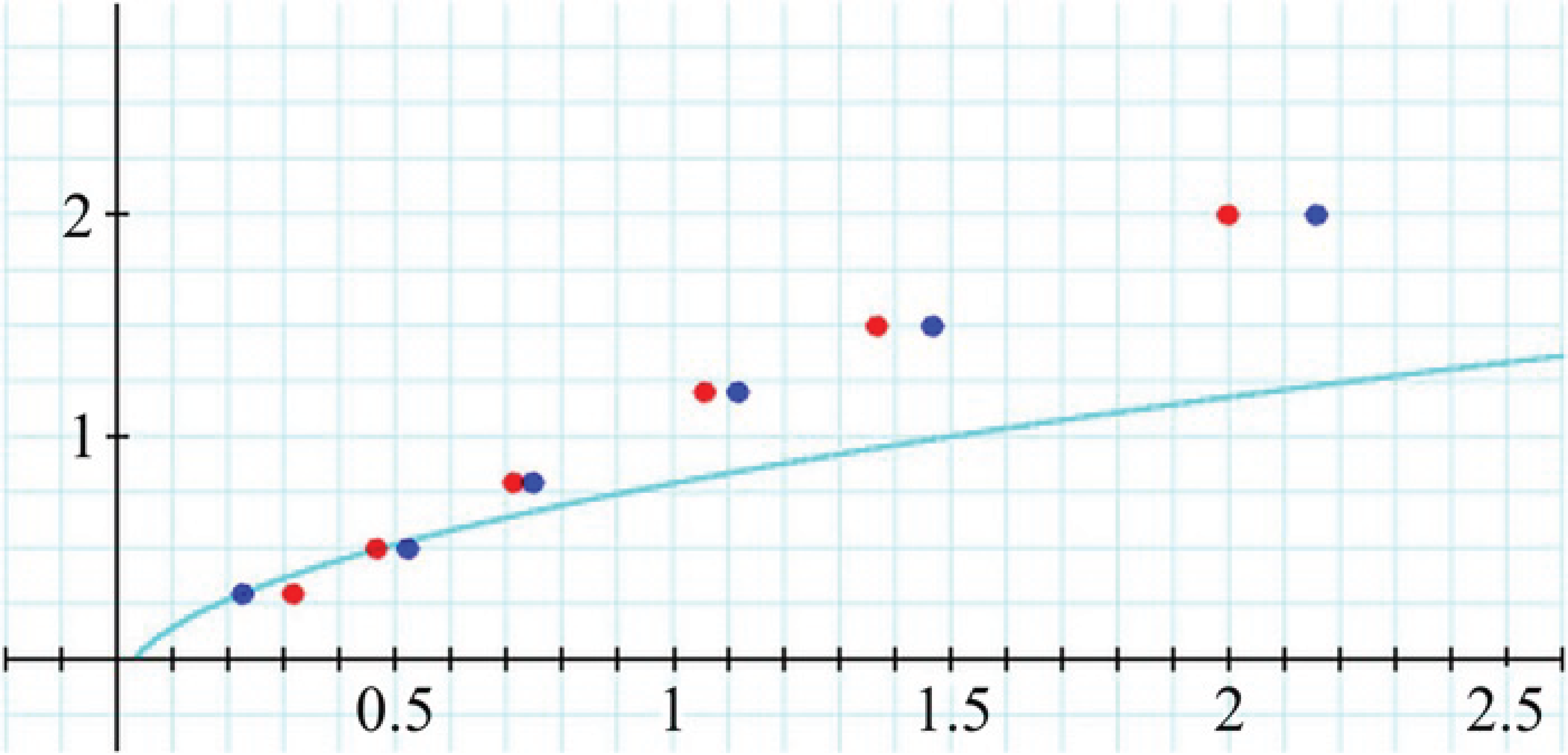}
    \caption{Speculative development of Venus and Pluto. The solid line is the ring-no-ring boundary. This would show why Venus and Pluto do not have rings.}
\end{figure}
  
%
%\epsfxsize.4\hsize
%\centerline{\epsfbox{Fig10.eps}}
%\vskip0cm
%\parindent=1pc
%\vskip1cm
%
% 
%
%\small{Fig.10. Speculative development of Venus and Pluto. The solid line is the ring-no-ring boundary. This would show why Venus and Pluto do not have rings.}
% 
 
While $ \eta \approx .1$ works well for most of the  planets, Venus and Pluto require a value closer to .01. In this figure both are plotted with this value of $\eta$. All planets except Saturn and Jupiter have ``trajectories'' determined by a best fit to the respective ellipsoid. 

\bb

\no{\bf \Large 5. Saturn and Jupiter. The Sun. Haumea }

The search for a simple model for Saturn, initially with no expectation of accounting for anything more than the equatorial bulge, revealed that
rings are a dominant feature of our model. The existence of rings gives us an additional measurable parameter, the mean radius standing in for the parameters of a complicated ring system. The radius and the width of the ring can be chosen within wide limits, but the model has not accounted for the flat ring system that is actually seen. The radius of a ring is closely related to the value of $\eta$.

In the case of Saturn, we began our study by looking for a fit to the main body, ignoring the extravagant system of rings. Attempts to fit the model to ellipsoids without rings,  with an equatorial bulge ratio of 1:10 failed.   We did not persist in this, because:  1. Observation of the gaseous giants does not favor our model of a phase transition at the surface.  2. The surface of a gas sphere is not well defined: experimental data usually refer to isobars. See for example Marsh (2017) [29] and  Lindal, Sweetnam and Esleman (1985) [30].
\bb

A photograph taken by NASA above the North Pole of Saturn, shows a hexagonal flow and 6 predicted positions for future whorls, Fig 11.  This phenomenon has been characterized as a Rossby wave (Marsh 2017 [31], Rossby 1939) [32].

 %%%%%%%%%%%
% FIGURES 11 & 12 %
 %%%%%%%%%%%
  
\begin{figure}[H]
    \centering
    \begin{minipage}{0.45\textwidth}
        \centering
        \includegraphics[width=\hsize]{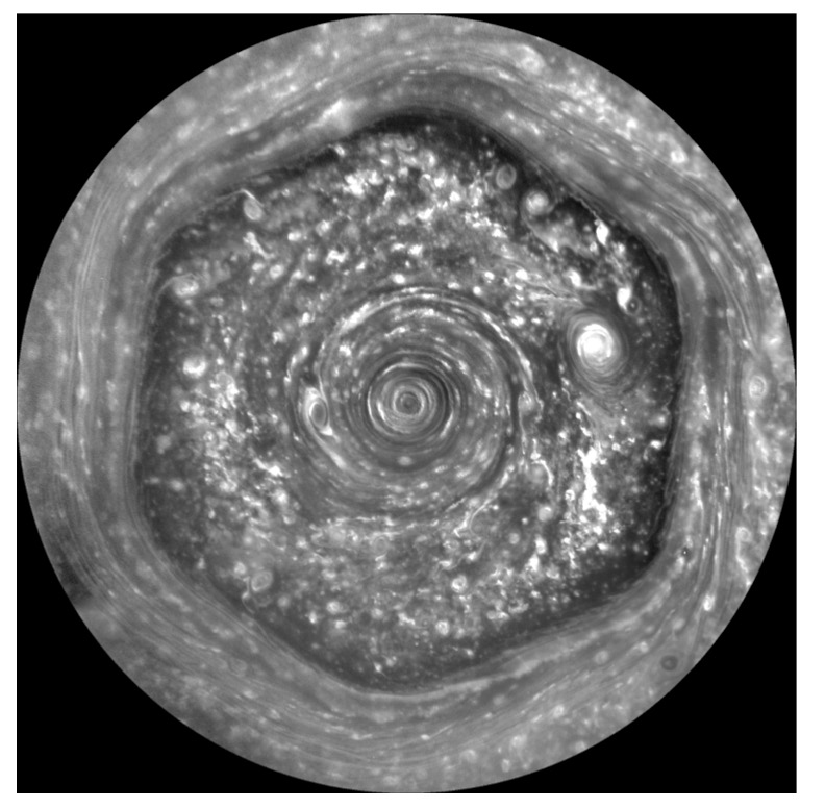} 
        \caption{This photograph taken by Voyager of the North pole of Saturn shows a regular hexagon flow as well as a remarkable, solitary, internal moon.  It shows a remarkable similarity to the shape with  six whorls of the symmetric model in the next figure.}
    \end{minipage}\hfill  
    \begin{minipage}{0.45\textwidth}
        \centering
        \includegraphics[width=\hsize]{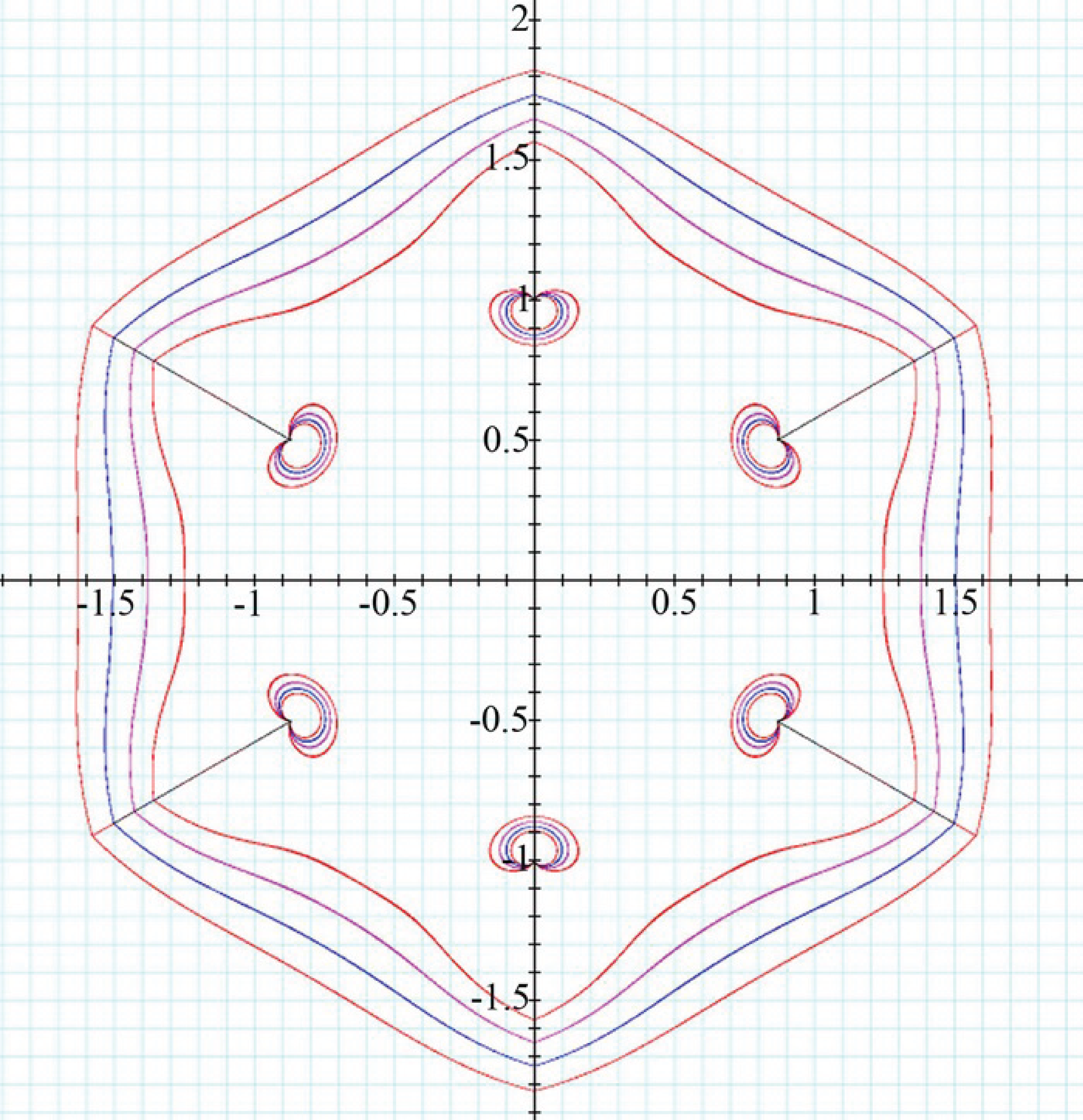}
        \caption{Flow lines of the function (5.2) with $a = k = 1, A = .5$.  This flow is a {\bf stationary solution} of our model;  a {\bf prediction} for future observations.}
    \end{minipage}
\end{figure}

%
%\b
% \vskip.5cm
%\epsfxsize.6\hsize
%\centerline{\epsfbox{Fig11.eps}}
%\vskip0cm
%
%\b
%
% 
%\epsfxsize.6\hsize
%\centerline{\epsfbox{Fig.11.eps}}
%\vskip0cm
%
%
%   
%\no Fig. 11. This photograph taken by Voyager of the North pole of Saturn shows a regular hexagon flow as well as a remarkable, solitary,  internal moon.  It shows a remarkable similarity to the shape with  six whorls of the symmetric model in the next figure.  
%
% 
%
%\epsfxsize.6\hsize
%\centerline{\epsfbox{Fig.12.eps}}
% \parindent=1pc
%
%\no \small{Fig. 12. Flow lines of the function (5.2) with $a = k = 1, A = .5$.  This flow is a {\bf stationary solution} of our model;  a {\bf prediction} for future obesrvations.
% \b

Refer to the general expression (2.4) for the velocity. To produce the picture  in Fig. 12 we assumed  a purely irrotational flow ($\tau = 0$) with 6-fold two-dimensional
rotational symmetry: 
$$
\Phi = \Im \bigg(a\ln z + A(1 +(kz)^{6})^{-1}\big)\bigg),~~~ A = {\rm constant.},....z:= x+iy. \eqno(5.1)
$$ 
The first term is the elementary flow  $a \DD\theta = (A/r^2)(-y,x,0)$ with unidirectional, circular  flow; the remainder has six-dimensional rotational symmetry.
    The gradient of $\Phi$  is
$$
d\Phi = 
$$
$$
\Bigg(a+{A(kr)^6\over D^2}\bigg(\big(1+(kr)^{12}\big)\cos(6\phi) + 2(kr)^6\bigg)\Bigg)d\phi
+ {A \sin(6\phi)\over D^2}(1-(kr)^{12}) (kr)^5 k dr.
$$
where $1/D = |1+(kz)^6|$.
For the square we get the surprisingly simple formula 
$$
(\DD\Phi)^2 = {a^2\over r^2} 
+{2a\over r^2}{A(kr)^6\over D^2}\bigg(\bigg(1+(kr)^{12})\cos(6\phi) + 2(kr)^{12}\bigg)
+  {(kA)^2(kr)^{10}\over D^2 }. 
$$
To  first order in the perturbation the modified expression for the function $f$ in Eq. (2.14) becomes, for some constant $\alpha$,
$$
 f(R,r):  
  = {(1 +  NR + \eta r^2)^2 - (N+1)^2 \over r^2} 
  $$
  $$
  + \alpha {r^4\over D^2}\bigg((1+(kr)^6)\cos(6\phi) +2(kr)^6\bigg)  + {\xi\over R}.
  $$

It turns out to be possible  to
produce hexagonal rings, but no further contact with observation was discovered, so far. However, with the extra term it becomes possible to imitate the ellipsoidal shape of Saturn.

The hexagon pattern on Saturn has been alternatively described as a Rossby wave.  [32].

Jupiter presents some of the same difficulties for analysis, but lacks the interesting hexagonal feature of Saturn. We have not constructed a model for the largest  planet.

%The Sun presents an anomaly; the period is about a day, but there is no apparent bulge. Is this a challenge? According to Eq.s(2.10) and (2.8) it implies that

The Sun is still further from our present objective, and so are galaxies.
We present, however, in Fig. 5.3,  an object that recalls, by its flatness, the shape of some galaxies;  the aspect ratio is about $10^9$.

 %%%%%%%%%%%%
% FIGURE THIRTEEN %
 %%%%%%%%%%%%
 
 \begin{figure}[H]
     \centering
     \includegraphics[width=0.6\hsize]{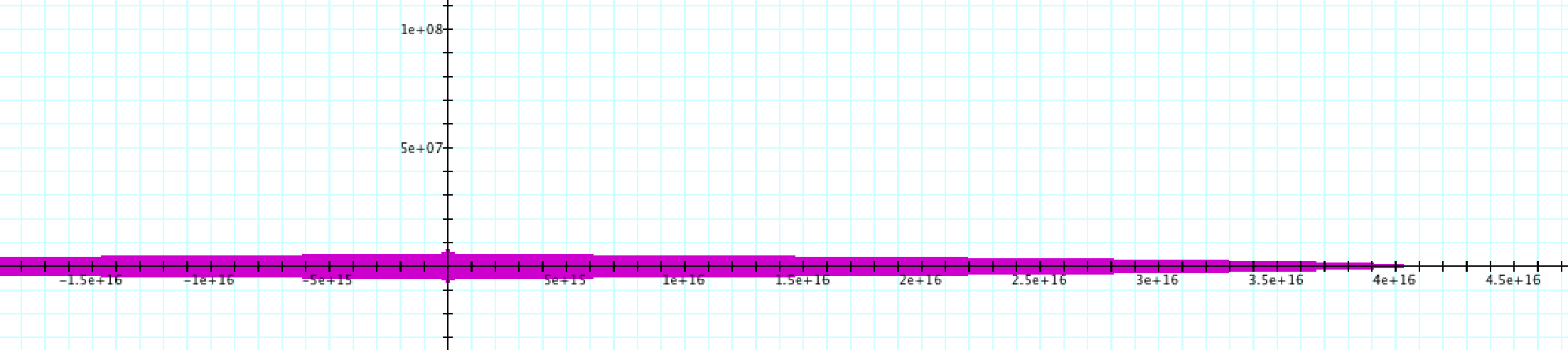}
     \caption{A shape produced by the model, suggesting an application to galaxies (cut off at the left)}
 \end{figure}
 
% \vskip0cm
% 
%\epsfxsize.6\hsize
%\centerline{\epsfbox{Fig13.eps}}
%\vskip0cm
%
%
%
%\parindent=1pc
%
%\small{Fig. 13. A shape produced by the model,  suggesting an application to galaxies
%(cut off at the left)}.  
%
%\b

Finally, here is a portrait  of one of the smallest object in the solar system, 
Haumea is a small moon or mini-planet in the outer Kuiper belt, remarkable  for its odd shape.

 %%%%%%%%%%%%%
% FIGURE FOURTEEN %
 %%%%%%%%%%%%%
 
 \begin{figure}[H]
     \centering
     \includegraphics[width=0.6\hsize]{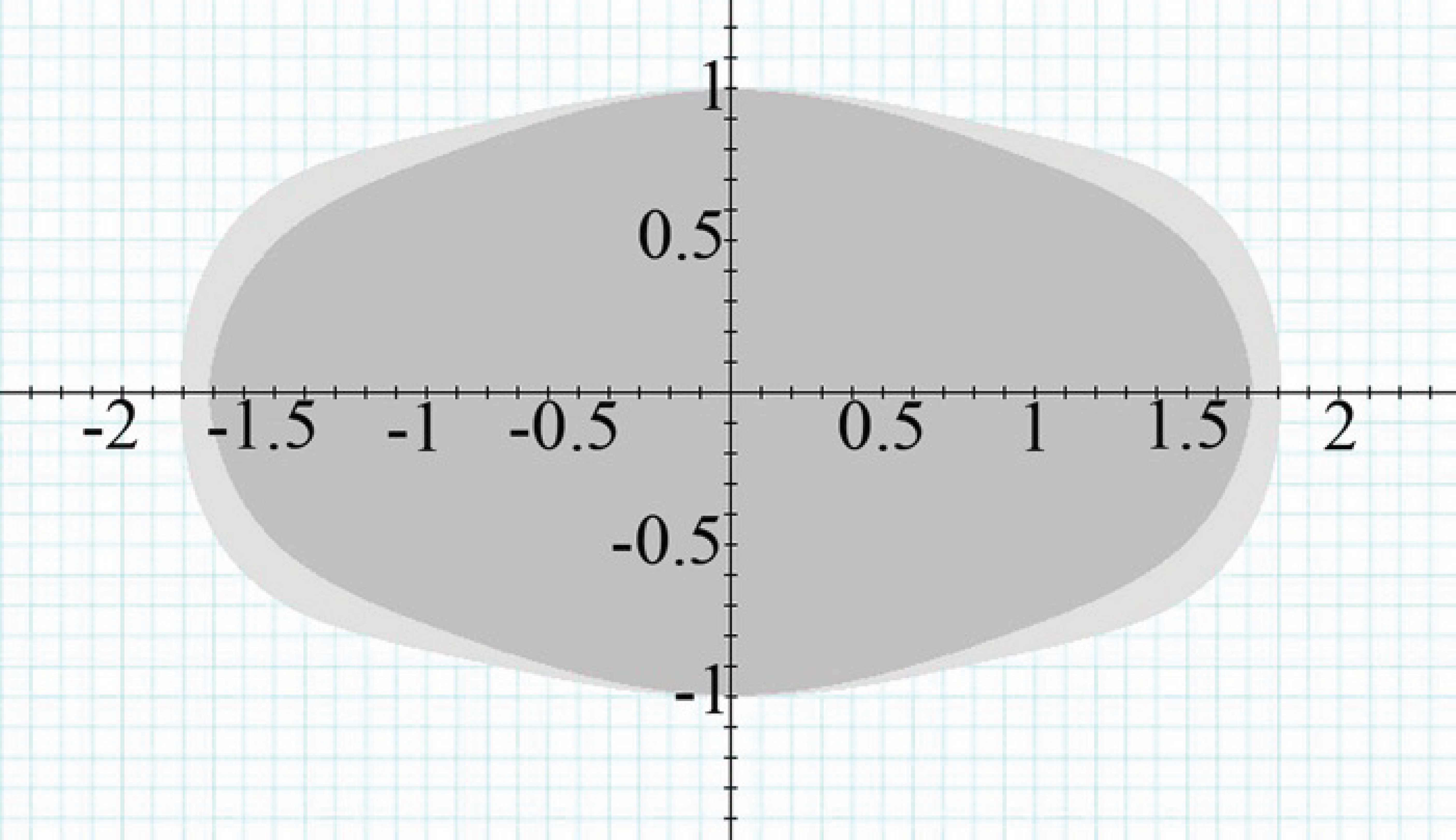}
     \caption{A model likeness to Haumea.}
 \end{figure}
 
 %\vskip1cm
 
%   \epsfxsize.4\hsize
%\centerline{\epsfbox{Fig14.eps}}
% \parindent=1pc
%
% \vskip1cm
%  
%\small{Fig. 14. A model likeness to Haumea.}

And here is the real Haumea.

%%%%%%%%%%%%
% FIGURE FIFTEEN %
%%%%%%%%%%%%
 
 \begin{figure}[H]
     \centering
     \includegraphics[width=0.6\hsize]{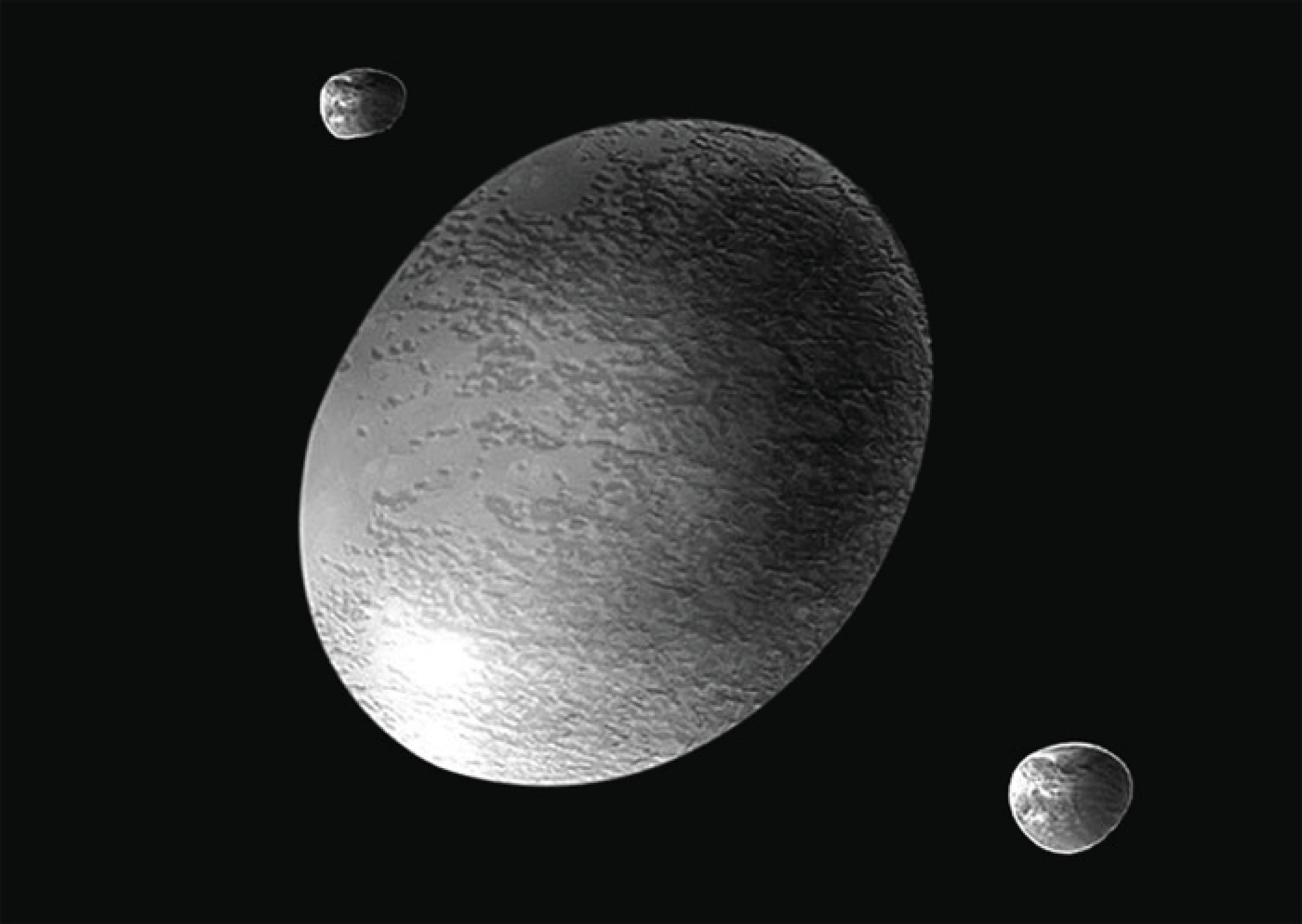}
     \caption{A photograph of Haumea taken by Voyager. The ring is barely visible on the original photograph.}
 \end{figure}

%\vskip1cm
%\hskip-1cm                     
% \epsfxsize.65\hsize
%\centerline{\epsfbox{Fig.15.eps}}
%\vskip0cm
%
%\parindent=1pc
%
%\vskip.5cm
% 
%
%\small{Fig. 15.  A photograph of Haumea taken by Voyager.  The ring is barely visible on the original photograph.}
%
%

\bb\no{\bf \Large 6. Summary and conclusions}

\b

The most significant result of this paper is that an action for hydrodynamics actually provides an effective  approach to real, astrophysical problems. The discovery that the rings seen on minor planets are natural within the formalism is a real surprise and source of encouragement. It is interesting to invent a special historical event sequence for each ring system, but we can now expect that some of them may be the result of a natural development.  
 
\b  

%\ce{\bf 6.1. The rings of Saturn}

All the models presented in this paper, except for the case of Saturn, postulate the simplest possible flows. The rings of the Jovian planets, with their system of multiple and very thin rings, present a challenge.  Figure (16)   shows a part of a model   with a very different flow,
with 
$$
\Phi = \tau =J_0(kr)\sinh(kz),~~~~\X=0. \eqno(6.1)
$$ 

%%%%%%%%%%%%
% FIGURE SIXTEEN %
%%%%%%%%%%%%
 
 \begin{figure}[H]
     \centering
     \includegraphics[width=0.6\hsize]{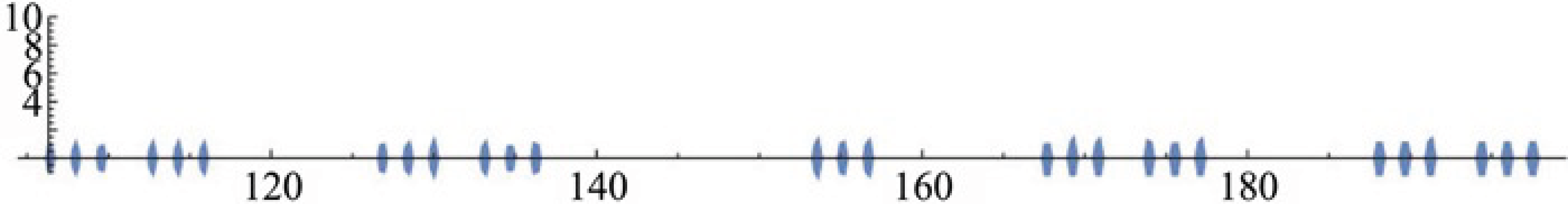}
     \caption{An attempt to imitate the rings of Saturn, showing a portion of a thin neighborhood of the equatorial plane, with a system of rings at semi-regular intervals. The parameter $k$ sets the scale, the same in both directions. The thickness of the rings is reduced by increasing $k$; in the illustration $k=2$.}
 \end{figure}
  
% 
%
% \vskip.5cm
%
%
%\vskip0cm
%\epsfxsize.6\hsize
%\centerline{\epsfbox{Fig.16.eps}}
%\vskip0cm
%
%\parindent=1pc
%
%\vskip.5cm
%
%   
%Fig.  16.
% An attempt to imitate the rings of Saturn, showing a portion of a thin neighborhood of the equatorial plane, with a system of rings at semi-regular intervals. The parameter $k$ sets the scale, the same in both directions. The thickness of the rings is reduced by increasing $k$;  in the illustration
%$k=2$.
%\b
%
% 
 
1. The equatorial bulge of a planet changes the gravitational field and this affects the calculations, though often to a minor degree. For each of the objects in the solar system it is
important to consider other complications that have hardly been mentioned in this paper, including the following.
\b

2.  The atmosphere was treated as empty; that leaves room for improvements. 
For the gaseous planets another model, without  a surface or with a discontinuous   density of the van der Waals type, may be indicated. The effect of magnetic fields and radiation must be included. We may attempt to extract an equation of state from the 
calculations in Section 2.3.  Without a determination of the entropy this can only amount to an evaluation of the value of the internal energy density. 
 If we could eliminate the coordinates from Eq.(2.13) in favor of the density, then that would give us the function $u(\rho,s)$. But since the left side of that equation depends on two of the coordinates this  would give a result that, at best, can be interpreted interms of a variable composition which is not envisaged by the model. However, this program cannot lead to a homogeneous equation of state.
\b
 
3. Angular momentum conservation  is a subtle issue that merits a separate
investigation. It is the sum of two parts, involving an orbital part $\vec x \w \vec v$ and the spin $\X \w  \vec{\ m}$. Only the sum is conserved but the first part is more likely to be observed. Even though the solutions found are stationary in the usual sense  we have found a 
linear dependence on $t$ in each of the two parts.
\b

4. A  problem that can be approached in the same spirit is the shape of galaxies. The inclusion of stress has a sensible effect on the famous anomalous velocity curves.

\bb

\no {\bf \Large Acknowledgements.}

An early version of this paper was posted on arXiv.org in May 2018 as arXiv 1803.09625 gen-ph and included in the book  (Fronsdal 2020a). Copyright Christian Fronsdal.
\b

Data available from the author.
 \bb
 
 \b 
  
{\bf \Large References} 

\no [1] \no Lagrange, J.M.,  Taurinensia, ii.,  Oeuvres, Paris, 1867-92 (1760)

 \no [2] Bernoulli, D.,``Hydrodynamica'' , Dulsecker, Argentoratum (1738)
 
 \no [3] D'Alembert, J., ``{\it Essay dune Nouvelle Theorie
 de la Resistance  des Fluides.} 
 
  Acad. R. des Sciences  
 de Paris (1747).
 
 \no \no[4] Fronsdal, C., {\it Adiabatic Thermodynamics of Fluids. From 
Hydrodynamics 

to General Relativity}. World Scientific (2020a) 
 
\no [5] Landau, L., ``Theory of  Superfluid  Helium II'', Phys. Rev.
 {\bf 60 }, 356-358 (1941)

\no [6] Rasetti, M., Regge, T.: ``Quantum vortices and diff (R3).'' In: 
{\it Lecture Notes in 

Physics}, Volume 20.  Physica {\bf 80 A}, 217 (1973)   

\no [7] Kalb, M. and Ramond, P. ``Classical direct interstring action". 

Phys. Rev. {\bf D 9} (8): 2273-2284 (1974)

\no [8]   Zheltukin, A., ``On brane symmetry'', arXiv.1409.6655. 

\no [9] Ogievetskij, V.I. and Palubarinov, V.  ``Minimal interactions between spin 0 and 

spin 1 fields",
J. Exptl. Theoret. Phys. (U.S.S.R.) {\bf 46} 1048-1055 (1964)

\no [10] Seliger, R.L. and Whitman, G.B, ``Variational principles in continuum mechanics'',

Proc.Roy. {\bf A 305}, 1-25 (1968)

\no [11]  Fronsdal, C., ``Stability analysis of Cylindrical Couette flow of compressible 

fluids, Phys. of Fluids'',  {\bf 32} 126117   doi 10.1063/5.0031200  {2020)

\no [12] Fronsdal, C., ``Hydrodynamical Sources for Gravitational Waves", in progress. 

\no [13] L. Beauvalet, L.,  Lainey, V, , Arlot, J.-E. and Binzel,  R. P.   

``Dynamical parameter determinations in Pluto's system, 

Expected constraints from the New Horizons mission to Pluto code'', 

A\&A 540, A65 
DOI: 10.1051/0004-6361/201116952 (2012)

\no [14] Kolb, S.M., Stute, M., Kley, W. and Mignone, A. ``Radiation hydrodynamics

integrated in the PLUTO 'code''. A\&A 559, A80 

DOI: 10.1051
/
0004-6361
/
201321499 (2013)

\no  [15] Schutz, B.F. Jr., ``Perfect fluids in General Relativity, Velocity potentials 

and a variational principle", Phys.Rev. {\bf D2}, 2762-2771 (1970)

\no  [16] Fronsdal, C. ``Ideal stars in General Relativity", 

Gen. Rel. Grav. {\bf 39}, 1971-2000 (2007)

\no [17]  Taub, A.H., ``General Relativistic Variational Principle for  Perfect Fluids", 

Phys. Rev. {\bf 94}, 1468  (1954)
     
 \no [18] Hall, H.E. and Vinen, W.F., ``The Rotation of Liquid Helium II. The Theory 
 
 of Mutual Friction in Uniformly Rotating Helium II",
Proc. R. Soc. Lond. A  {\bf }23}, 

doi: 10.1098/rspa.1956.0215 (1956)  

\no [19 Fetter A.I., ``Rotating Trapped Bose-Einstein condensates'',
  Rev. Mod. Phys. {\bf 81} 647 
  
  (2009)

\no  [20] Lane, J.H., ``On the Theoretical temperature of the sun; under the hypothesis 

of a Gaseous Mass maintaining its Volume by its internal Heat, and depending

on the Laws of Gases as known to Terrestrial experiment'',

Am.J.Sc.\&Arts, {\bf 50} 57-74 (1870)

\no [21] Couette, M., ``Oscillations tournantes d'un solide
de r\'evolution en contact 

avec un fluide  visqueux,'' C. R. Acad. Sci. Paris
{\bf 105}, 1064-1067 (1887) 

\no [22] Mallock, A., Proc.R.Soc. {\bf 45} 126 (1889)

\no [23] Mallock, A., Ohilos.Trans.R.Soc. {\bf 187}, 41 (1896) 

\no [24]  Fitzpatrick, R. ``Newtonian Mechanics", Lecture notes, University of Texas, 

available on  Prof. Fitzpatrick's web page (2020) 

\no [25]  Carr, M. and Head, {\it  The geological history of Mars,} 2010

\no  [26] Carr, M. (2006). {\it {\it} The surface of Mars.} Cambridge, UK: 

Cambridge University Press. ISBN 0-521-87201-4  (2006) 

\no  [27] Bertka, C.M. and Fei, Y., ``Density profile of an SNC model Martian interior 

and the     25-moment-of-inertia factor of Mars'',   
 EP Sci.Lett. {\bf 157} 79-88 (1990)
 
\no [28] Helled, R., Anderson, J.D. and Schubert, G. ``Uranus and Neptune:

Shape and rotation." Icarus {\bf 210} 446-454 (2010) 
  
\no  [29] Marsh, G. E., ``Enigma of Saturn’s North-Polar Hexagon”, PACS,

96.30.Mh 96.15 Hy; 96.15.Xy  arXiv:1711.00338 17 [physics.ao-ph]  (2017) 

\no [30] Lindal, G.F., Sweetnam, D.N. and Esleman, V.R. ``The atmosphere of saturn: an 

analysis of the voyager radio occultation measurements". 

The Astronomical  J. {\bf 90} (6) 1136-1147, June (1985)

\no [31] Rossby, C.-G., 1939: ``Relation between variations in the intensity of the zonal 

circulation of the atmosphere and the displacements of the semi-permanent centers of 

action.", J.  Mar. Res., {\bf 2}, 38–55  (1939) 

\ve

\no{\bf \Large Appendix 2. Tables}

 \begin{table}[h]
\caption{The solar system}
\begin{center}
\begin{tabular}{@{}cccccccc@{}}

&\hskip0cm  &   Radius & Mass & Density & Bulge  &   Period &  
\cr

&\hskip0cm Body  & ~Km &$10^{23}$ g& $g cm^{-3}$ &~ratio&    ~Days &  
\cr
\hline 
&\hskip0cm Moon & 1738 & 735 &  3.34 & small & 28 &  
\cr

&\hskip0cm  Pluto & 1188 & 318E & 2.03 & 1:3 300   &    1.4
\cr

&\hskip0cm Mercury & 2,440 & 3,300 & 5.4 & small   & 59 & 
\cr

&\hskip0cm  Mars &3376 & 6,419 & 3.93 & 1:136 & 1.026 &   
\cr

 &\hskip0cm  Venus & 6500 & 5680 & 5.2 &1:3,200 & 243  &     
\cr

&\hskip0cm  Earth & 6,357 & 60,000 &  5,515 & 1:300   & 1 &  
\cr

&\hskip0cm Uranus & 24,973 & 8.66E+5 & 1.318 & 1:44&370
 &  
\cr

&\hskip0cm Neptune  & 34,342  & 1.03E+6 & 1.638  & 1:59   &367 
\cr

&\hskip0cm  Saturn & 54,364 & 5.69E+6 & .687 &1:10   &    .44 & 
\cr

&\hskip0cm  Jupiter &66,854 & 1.9E+7 & 1.326 & 1:14   &    .414 & 
\cr

&\hskip0cm  Jupiter &66,854 & 1.9E+7 & 1.326 & 1:14   &    .14
\cr

&\hskip0cm  Sun & .7E+6 & 2.0E+10 & 1.41 & 0   & 24&   
\cr
\hline
\end{tabular}
\end{center}
\end{table}

\bb
 \begin{table}[h]
\caption{The parameters}
\begin{center}
\begin{tabular}{@{}cccccccccccc@{}}

%\settabs \+ & \hskip1.5cm    & \hskip1.5cm  & \hskip1.5cm  &\hskip1.5cm &\hskip1.5cm & \hskip1.5cm & hskip1.5cm&hskip1.5cm&\hskip1.5cm
%\cr
    & Body &  Radius & 2GM & Period & N & $\xi$ & $\sigma$ & $\kappa$  
\cr 
%\vskip1mm\hrule\vskip1mm        
&\hskip0cm   Earth &  1  & 1 & 1 & 1.5 & 3.4 & 2.8 &  8.2
\cr        
&\hskip0cm  Mars & .53 &  .11 & 1.03  & .8  & 1.5 & 2.4 & 2.1
\cr       
&    Neptune & 3.9 & 17.2 & .67 & .2 & .68 & 4.0  &  27
\cr
&\hskip0cm   Uranus &4.01 & 14.6 & .72 & .2 & .55 & 4.2  & 33
\cr                 
&\hskip0cm   Venus & .95 & .82 & 243 & .3 & .23 &  9.6 & 7380
\cr        
\hline
\end{tabular}
\end{center}
\end{table}

\bb\b

\end{document}